\documentclass[aps,prb,reprint,showpacs,twocolumn,floatfix]{revtex4-2}

\usepackage{graphicx}
\usepackage{bm}
\usepackage[utf8]{inputenc}
\usepackage[T1]{fontenc}
\usepackage{amsmath}
\usepackage{amssymb} 
\usepackage[capitalise]{cleveref}

\renewcommand{\arraystretch}{0.7}

\begin{document}

\title{\texorpdfstring{\textit{Ab initio} design of quaternary Heusler compounds for reconfigurable magnetic tunnel diodes and transistors}{Ab initio design of quaternary Heusler compounds for reconfigurable magnetic tunnel diodes and transistors}}

\author{T. Aull$^{1}$}
\author{E. \c{S}a\c{s}{\i}o\u{g}lu$^{1}$}
\author{I. V. Maznichenko$^{1}$}
\author{S. Ostanin$^{1}$}
\author{A. Ernst$^{2,3}$}
\author{I. Mertig$^{1,2}$}
\author{I. Galanakis$^{4}$}

\affiliation{$^{1}$Institute of Physics, Martin Luther University Halle-Wittenberg, D-06120 Halle (Saale), Germany \\
$^{2}$Max Planck Institute of Microstructure Physics, Weinberg 2, D-06120 Halle (Saale), Germany\\
$^{3}$Institute for Theoretical Physics, Johannes Kepler University Linz, Altenberger Straße 69, A-4040 Linz, Austria\\
$^{4}$Department of Materials Science, School of Natural Sciences, University of Patras, GR-26504 Patra, Greece}



\begin{abstract}

Reconfigurable magnetic tunnel diodes and transistors are a new concept in spintronics. The realization of such a device requires the use of materials with unique spin-dependent electronic properties such as half-metallic magnets (HMMs) and spin-gapless semiconductors (SGSs). Quaternary Heusler compounds offer a unique platform to design within the same family of compounds HMMs and SGSs with similar lattice constants to make coherent growth of the consecutive spacers of the device possible. Employing state-of-the-art first-principles calculations, we scan the quaternary Heusler compounds and identify suitable candidates for these spintronic devices combining the desirable properties: (i) HMMs with sizable energy gap or SGSs with spin gaps both below and above the Fermi level, (ii) high Curie temperature, (iii) convex hull energy distance less than 0.20 eV, and (iv) negative formation energies. Our results pave the way for the experimental realization of the proposed magnetic tunnel diodes and transistors.

\end{abstract}

\maketitle

\section{Introduction \label{sec:In}}

The growing interest in nanotechnology in the last decades laid the foundation of research in new materials with novel properties. In particular, the prediction of new magnetic nanomaterials for the realization of spintronic devices has become extremely important~\cite{vzutic2004spintronics}. There are two ways to incorporate spin in electronic devices: either doping semiconductors with magnetic ions like Mn, Cr, or Fe in diluted magnetic semiconductors~\cite{sato2010first} or the growth of nanoscale magnetic materials like Heusler compounds~\cite{katsnelson2008half}. 
The development of computational materials science triggered all these developments in spintronics. In particular, computational materials science paved the way for high-throughput screenings, which permitted efficient simulations of materials in order to predict magnetic, optical, and electronic characteristics, etc., of new materials. Furthermore, the simulations allowed researchers to investigate new metastable structures of known alloys where the electronic features change completely concerning the properties of the known stable structures.
Among the various materials under study for spintronics and magnetoelectronics, magnetic Heusler compounds have a significant importance due to their wide variety and their high Curie temperatures, and thus several studies covering their fundamental properties and their applications have been carried out~\cite{Perspectives}. Among the magnetic Heusler compounds, several have been identified as half-metallic magnets \cite{Ma,Ma2,Sanvito,Faleev,Faleev2}. Also, even more peculiar properties have been suggested in literature like spin-gapless semiconducting or spin-filtering properties, which lead to new functionalities~\cite{GalanakisAIP}. Modern deposition techniques made fabrication of these exotic materials possible. A recent example is (CrV)TiAl, a quaternary Heusler compound which was predicted in Ref.~\cite{Galanakis14} to be a fully compensated ferrimagnetic semiconductor, and then it was grown successfully and its unique magnetic properties have been confirmed~\cite{Venkateswara}. Thus, there is merit in the study of this family of alloys and compounds.

\begin{figure}[t]
\begin{center}
    \includegraphics[width=0.49\textwidth]{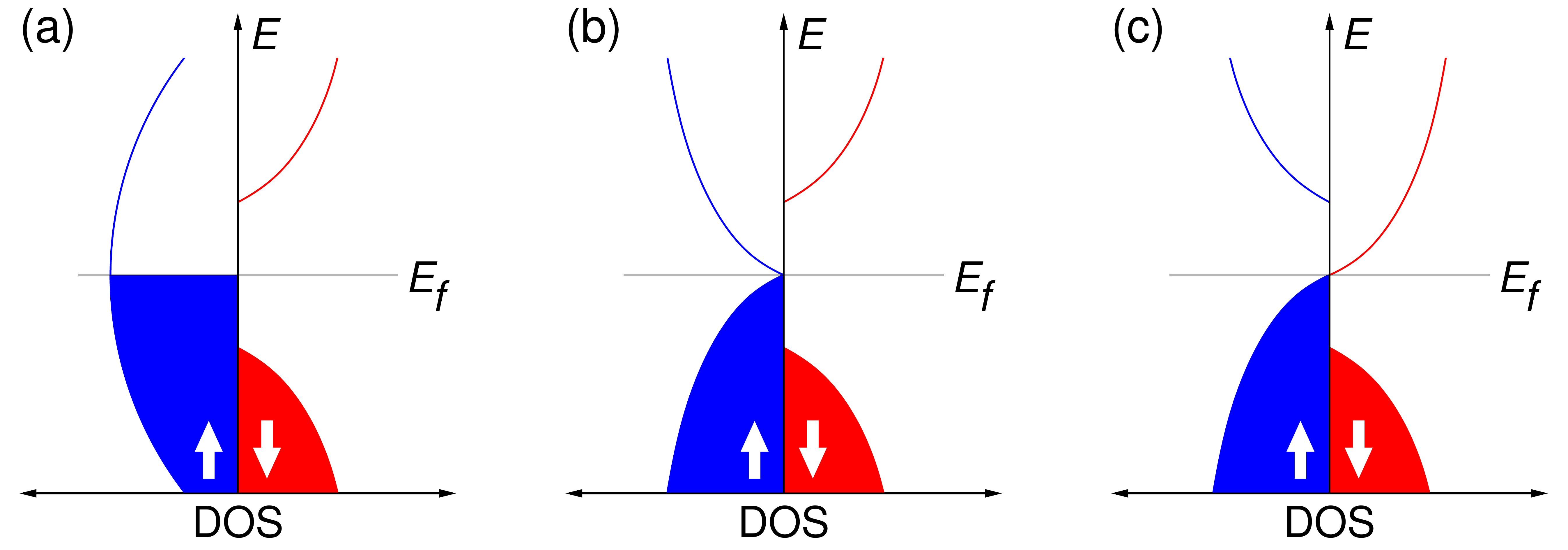}
\end{center}
\vspace{-0.5cm}
\caption{Schematic representation of the density of states (a) for a half-metallic magnet (b) for a type-I spin-gapless semiconductor, and (c) for a type-II spin-gapless semiconductor.}
\label{fig:SGS-HMM}
\end{figure}

A special class of materials, mentioned above, receiving substantial interest is the so-called gapless semiconductors, in which conduction- and valence-band edges touch at the Fermi level~\cite{tsidilkovski2012electron}. In such materials, the mobility of charge carriers is essentially much higher than in normal semiconductors, making them promising materials for nanoelectronic applications. The first gapless semiconductors that have been studied were Hg-based IV-VI compounds, especially HgCdTe, HgCdSe, and HgZnSe. But it turned out that all these alloys are toxic and oxidize easily~\cite{tsidilkovski2012electron}. 
Later, Kurzman \textit{et al.} proposed PbPdO$_2$ as a gapless semiconductor~\cite{kurzman2011hybrid} 
and its zero gap width was demonstrated experimentally~\cite{chen2011gapless}. Nowadays, 
one of the most studied gapless semiconductors is graphene~\cite{novoselov2004electric}. 
In 2008 Wang proposed that doping PbPdO$_2$ with Co atoms would result in a new class of 
materials: the so-called spin-gapless semiconductors (SGSs) (see Ref.~\cite{wang2008proposal,wang2009colossal}). 
The spin-gapless semiconductors lie on the border between half-metallic magnets 
(HMMs)~\cite{de1983new} and magnetic semiconductors. A schematic density of states (DOS) of a HMM and a SGS (type I and type II) is shown in \cref{fig:SGS-HMM}. 
The spin-up (majority- spin) band in HMMs crosses the Fermi level like in a normal magnetic metal, but, in contrast to metals, in the spin-down (minority-spin) band a gap appears and the Fermi level lies in between the gap like in normal semiconductors. 
For type-I SGSs, the minority-spin band looks like in HMMs but the difference is in the majority-spin band. The valence- and conduction-band edges are touching at the Fermi energy so that there appears a zero-width gap. On the other hand, type-II SGSs possess a unique electronic band structure that there is a finite gap just above and below the Fermi energy $E_F$ for different spin channels, i.e., conduction- and valence-band edges of the different spin channels touch. Ferromagnetism is also possible in SGSs since the two spin band structures are different. One important advantage of type-I SGSs is that no energy is required for the excitation of the electrons from the valence to the conduction band and excited electrons or holes can be 100\% spin-polarized. 
It is worth noting that for type-II SGSs the spin-gapless semiconducting properties are not protected by any symmetry and can only appear if a free parameter, e.g., pressure, is tuned to a specific value.

Since the first proposal of spin-gapless semiconducting
properties in Co-doped PbPdO$_2$,
different classes of materials
ranging from three to two dimensions have been predicted
to possess SGS characteristics and a few of them have been
confirmed experimentally. Among them, graphene nanoribbons altered by CH$_2$ radical groups~\cite{pan2011exploration}, in which magnetism
originates from the unsaturated carbon states, show spin-gapless characteristics. HgCr$_2$Se$_4$ has a phase 
transition under a pressure of 9\,GPa from the ferromagnetic semiconductor to the SGS 
state~\cite{guo2012density}. The boron nitride nanoribbons with
vacancies present SGS properties~\cite{pan2010electronic}. \textit{Ab-initio} calculations from different groups have shown that several Heusler
compounds present either type-I or type-II SGS properties. 
Mn$_2$CoAl was the first Heusler compound, the type-I SGS
characteristics of which were experimentally demonstrated by
Ouardi \textit{et al.}~\cite{ouardi2013realization}. Furthermore, Mn$_2$CoAl 
possesses a high Curie temperature of 720 K~\cite{ouardi2013realization} and high electron 
and hole mobility. The search for SGSs has been extended recently to
the family of ordered quaternary Heusler compounds which
are usually named as LiMgPdSn-type Heuslers (also known
as LiMgPdSb-type Heusler compounds)~\cite{ozdougan2013slater,Xu2013}. They have the chemical 
formula (\textit{XX'})\textit{YZ} with transition-metal atoms \textit{X}, \textit{X'}, and \textit{Y}, where the valence of \textit{X'} is lower 
than the valence of \textit{X} atoms and the valence of the \textit{Y} element is lower than the valence of both 
\textit{X} and \textit{X'}. For reasons of simplicity usually in literature the parentheses are omitted and they are denoted as \textit{XX'YZ}. In 2013, two extended \textit{ab initio} studies have appeared focusing on their electronic and magnetic properties and several have been found to be 
SGSs~\cite{ozdougan2013slater,Xu2013}. Very recently, Gao \textit{et al.}, performed a 
systematic screening of the SGSs in ordered quaternary Heusler alloys focusing on the mechanical and dynamical stability and identified 70 stable SGSs demonstrating that four types of SGSs can be realized based on the spin characteristics of the bands around the Fermi level~\cite{gao2019high}.

\section{Motivation and aim \label{sec:Motiv}}

Spintronics and magnetoelectronics are two rapidly emerging fields in current nanoelectronics. HMMs have been considered as ideal electrode materials in magnetic tunnel junctions for spin-transfer torque magnetic memory applications due to their 100\% spin polarization of the conduction electrons at the Fermi level, which leads to a very high tunnel magnetoresistance (TMR) effect. Half-metallic Heusler com- pounds have been used by several experimental groups to fabricate magnetic tunnel junctions due to their very high Curie temperatures and lattice parameter matching with the conventional tunnel barrier MgO. High TMR effects have been experimentally demonstrated in tunnel junctions made of Co-based Heusler compounds~\cite{TMR_H_1,TMR_H_2,TMR_H_3}.

Although magnetic tunnel junctions made of half metals show large TMR effects making them very suitable for memory applications, they do not present any rectification (or diode effect) for logic operations. Logic-in-memory computing is an emerging field that promises to solve the bandwidth bottleneck issues in today’s microprocessors. In semiconductor nanoelectronic devices, despite intensive efforts, the combination of nonvolatility and reconfigurability on the diode (transistor) level has not yet been achieved. Recently this became possible by utilizing the unique spin-dependent transport properties of SGSs and thus a new spintronic device concept has been proposed in Ref.~\cite{sasioglu2015patent}, 
which combines reconfigurability and nonvolatility on the diode and transistor level. Furthermore, the proposed transistor overcomes the limitations of conventional hot electron quantum tunnel devices such as base-collector leakage currents in tunnel transistors~\cite{GBT_5}, 
which might lead to high power dissipation.

\begin{figure}[t]
\vspace{-0.3cm}
\centering
\includegraphics[width=0.45\textwidth]{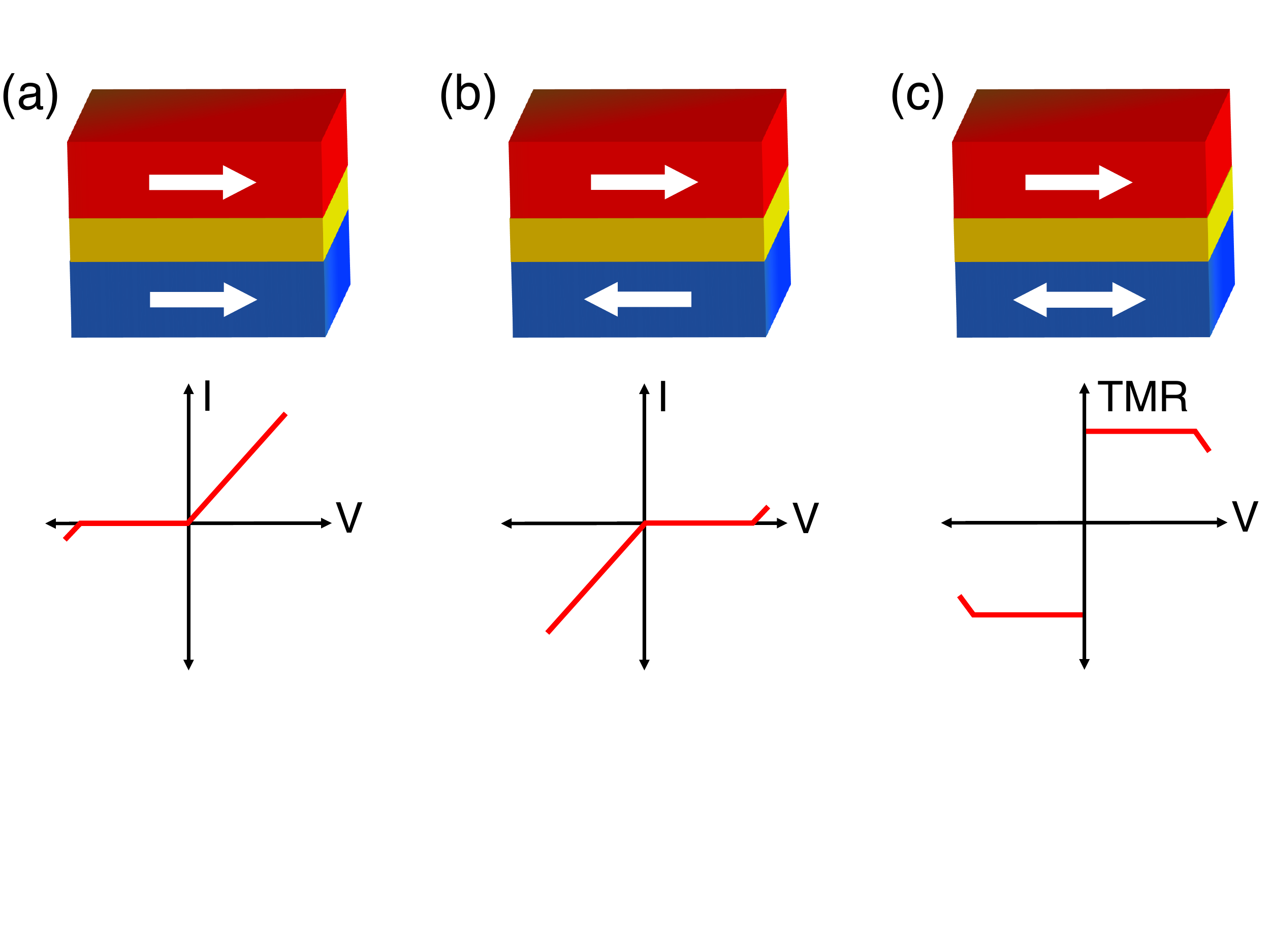}
\vspace{-1.9cm}
\caption{Schematic representation of the reconfigurable magnetic tunnel diode for (a) parallel and (b) antiparallel orientations of the magnetization directions of the electrodes and corresponding current-voltage (I-V) characteristics. (c) Bias voltage dependence of the TMR effect in a magnetic tunnel diode. With arrows we show the magnetization direction of the electrodes.}
\label{fig2}
\end{figure}

The principles of the proposed reconfigurable magnetic tunnel diode (MTD) and transistor (MTT) have been extensively discussed in a very recent article (see Ref.~\cite{sasioglu2019proposal})
and thus here we will present only a short overview of the proposed devices. The structure of the proposed reconfigurable MTD and its current-voltage (\textit{I-V}) characteristics are schematically shown in Fig.~\ref{fig2}. 
The MTD consists of a type-II SGS electrode and a HMM electrode separated by a thin insulating tunnel barrier and the rectification properties of the MTD are determined by the relative orientation of the magnetization directions of the electrodes. Using a type-I SGS instead of the HMM is also possible. When the magnetization directions of the electrodes are parallel to each other [see Fig.\,\hyperref[fig2]{2(a)}] then the tunneling current is only allowed in one direction; in the reverse direction the tunneling current is blocked. Thus, the tunnel junction behaves like a rectifier, i.e., a diode. When the magnetization direction of one of the electrodes is reversed, then the rectification properties of the diode are also reversed as shown in Fig.~\hyperref[fig2]{2(b)}. Hence, the MTD can be configured dynamically by current-induced spin-transfer torque or by an external magnetic field.

The first theoretical study on SGSs with type-II band structure within the Heusler family has been reported by two of the present authors in Ref.~\cite{ozdougan2013slater}. MTTs are an extension 
of the concept of MTDs where two back-to-back MTDs are used to build a three-terminal device as described in Ref.~\cite{sasioglu2019proposal}. The value of the gap in one spin channel for the HMMs and type-I SGSs as well as the gaps in different spin channels of type-II SGSs play a decisive role in determining the \textit{I-V} characteristics of the MTD as discussed in Ref.~\cite{sasioglu2019proposal}. 
Suitable SGSs and HMMs should have similar lattice constants so that the coherent growth of the device is possible. They should have high Curie temperatures, $T_C$, in order to be operational at room temperature. HMMs should possess large minority-spin gaps and SGSs should possess sizable gaps both below and above the Fermi level (for this reason, type-III and type-IV SGSs described in Ref.~\cite{gao2019high} are not suitable for such devices). And finally, in addition to negative formation energies, they should have a reasonably small convex hull energy distance so that their growth as metastable phases in the form of thin films could be feasible.

The aim of the present paper is to screen Heusler-based electrode materials with $T_C$ values above room temperature for realization of the new device concept. Especially for type-II SGSs, to the best of our knowledge, up to now neither theoretical nor experimental work has been reported addressing the finite-temperature properties contrary to type-I SGSs~\cite{ouardi2013realization,jakobsson2015first}. 
To this end, we focus on the HMMs and SGSs (type I and type II) in ordered quaternary Heusler structure \textit{XX'YZ}. In total, we identify 25 materials with sizable band gaps around the Fermi level which are either HMM or SGS and which fulfill the conditions mentioned above. In particular, for the SGS type-II materials, the tunability of the relative position of the valence-band maximum (VBM) and the conduction- band minimum (CBM) with substitution of different Z atoms is discussed. To study finite-temperature properties, we map the multisublattice complex itinerant electron problem onto the classical Heisenberg model with exchange parameters calculated using the Liechtenstein formalism~\cite{liechtenstein1987local}. 
e find that in agreement with previous studies due to the presence of a spin gap in both HMMs and SGSs the exchange interactions decay quickly with distance, and hence magnetism of these materials can be described considering only nearest- and next-nearest-neighbor intersublattice and intrasublattice exchange interactions. For all SGSs and most of the HMMs, the estimated Curie temperatures are above room temperature,
making them suitable candidates as electrode materials for
reconfigurable device applications. Furthermore, we show that
the $T_C$ values obey a semiempirical relation $T_C \sim \sum_i |m_i|$, i.e., $T_C$ increases with increasing sublattice magnetic moments. The rest of the paper is organized as follows. In
 Sec.~\ref{sec:CompDet} we describe the computational method while in Sec.~\ref{sec:resanddis} 
our results are presented and discussed. Finally, we summarize and present our conclusions in Sec.~\ref{sec:concl}.

\begin{figure}[t]
\centering
\includegraphics[width=0.45\textwidth]{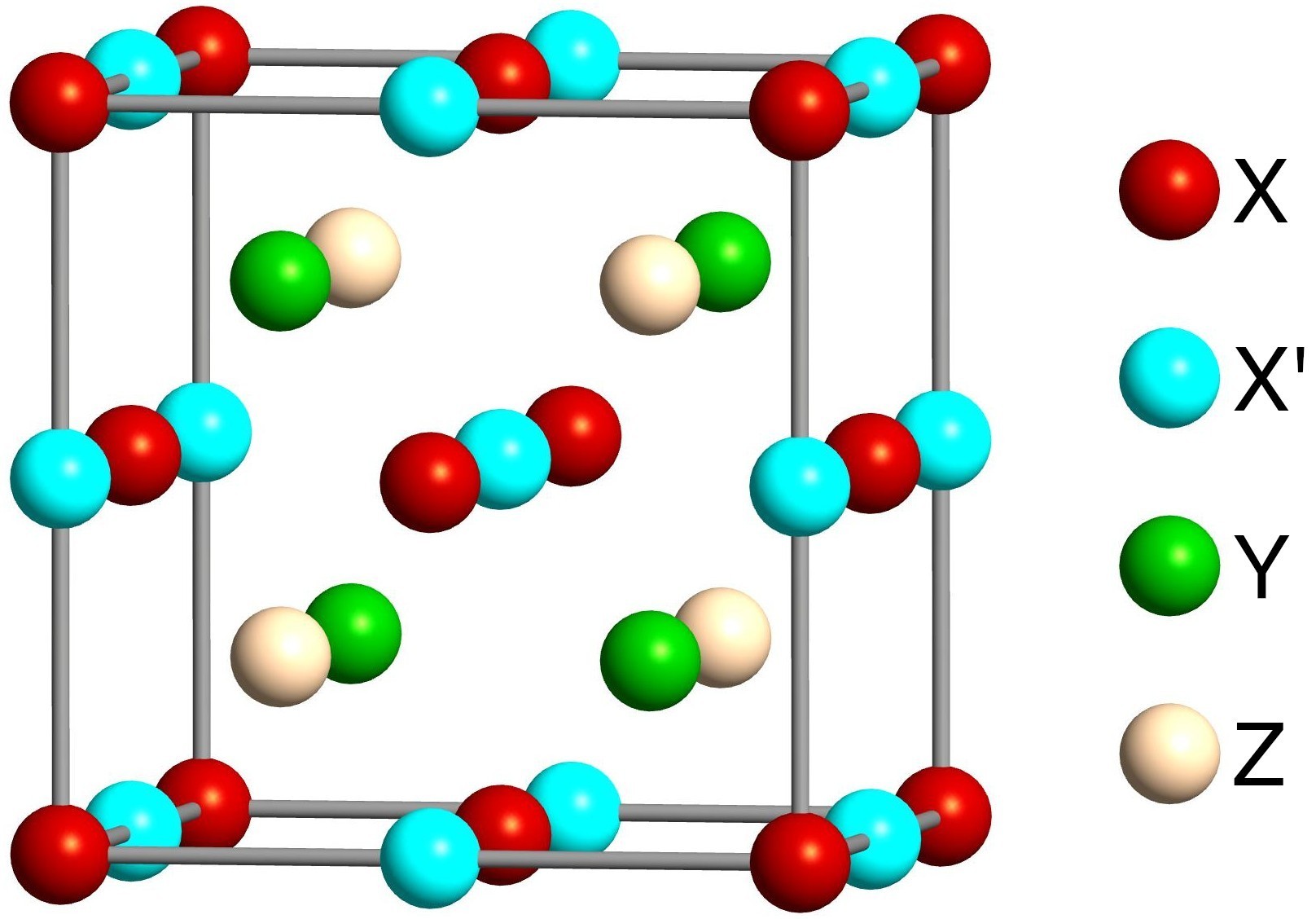}
\caption{Crystal structure of the quaternary Heusler alloys \textit{XX'YZ}.
\textit{X} is located at Wyckoff position 4\textit{a}(0,0,0), 
\textit{Y} is located at 4\textit{c}($\frac{1}{4}$,$\frac{1}{4}$,$\frac{1}{4}$), 
\textit{X'} is located at 4\textit{b}($\frac{1}{2}$,$\frac{1}{2}$,$\frac{1}{2}$) and
\textit{Z} is located at 4\textit{d}($\frac{3}{4}$,$\frac{3}{4}$,$\frac{3}{4}$).}
\label{fig:Heusler}
\end{figure}

\section{Computational Method \label{sec:CompDet}}

For all calculations, we consider Heusler compounds
with the chemical formula \textit{XX'YZ}. As mentioned above
\textit{X} , \textit{X'} , and \textit{Y} are transition-metal atoms with descending
valence and \textit{Z} is a metalloid. Ordered quaternary Heusler
compounds adopt the so-called LiMgPdSn-type cubic
structure with space group \textit{F}$\overline{4}$\textit{3m} (space group 216) (see \cref{fig:Heusler}), where the \textit{X} atoms 
occupy Wyckoff position 4\textit{a}(0,0,0), \textit{X'} 4\textit{b}($\frac{1}{2}$,$\frac{1}{2}$,$\frac{1}{2}$), \textit{Y} 
4\textit{c}($\frac{1}{4}$,$\frac{1}{4}$,$\frac{1}{4}$) 
and \textit{Z} 4\textit{d}($\frac{3}{4}$,$\frac{3}{4}$,$\frac{3}{4}$)~\cite{klaer2011element,wei2015electronic}. 
We should note that the \textit{X} and \textit{X'} atoms at 4\textit{a} and 4\textit{b} sites form a cubic lattice. The same is true for the \textit{Y} and \textit{Z} atoms sitting at the 4\textit{c} and 4\textit{d} sites. Overall the structure can be considered as fcc with four atoms as the basis along the long diagonal of the cube shown in Fig.~\ref{fig:Heusler} 
with the sequence \textit{X-Y-X'-Z}. Note that this occupation scheme of the elements is energetically the most favorable with respect to any exchange of the atoms at the various sites~\cite{gao2019high}.
The density functional theory (DFT) calculations were performed using the QuantumATK package~\cite{QuantumATK}, version O-2018.06, together with the norm-conserving 
\textsc{PseudoDojo} pseudopotentials~\cite{QuantumATKPseudoDojo}. 
We should note here that a recent study on SGSs using the GW approximation for the electronic self-energy to account for many-body exchange-correlation effects has shown that the effect of employing GW is small in the case of SGSs and the usual density functional theory gives a fair description of the electronic properties of these materials~\cite{Murat}. In the case of HMMs, the changes in the electronic structure by using GW should be even smaller due to their metallic character. For electronic structure calculations, we
used a linear combination of atomic orbitals (LCAO) method within the Perdew-Burke-Ernzerhof parametrization of the generalized gradient approximation functional~\cite{perdew1996generalized}
utilizing a $15 \times 15 \times 15$ Monkhorst-Pack grid~\cite{monkhorst1976special} and a density mesh cutoff of 120 hartree. The total energy and forces have been converged at least to 10$^{-4}$\,eV and 0.01\,eV/{\AA}, 
respectively. Since we are only discussing magnetic materials, all calculations were performed taking spin polarization into account with collinear aligned spins. We used the calculated equilibrium lattice constant for each material. Note that all considered materials are mechanically as well as dynamically stable~\cite{gao2019high}.\\

To study finite-temperature properties we map the complex multisublattice itinerant electron problem onto a classical effective Heisenberg Hamiltonian
\begin{equation}
\label{eqn:Heisenberg}
    H_{\mathrm{eff}} = -\sum_{\mathrm{i,j}}\sum_{\mathrm{\mu,\nu}} J^{\mu\nu}_{\mathrm{ij}} 
    \bm S^{\mu}_i \cdot \bm S^{\nu}_j, 
\end{equation}
where $\mu$ and $\nu$ denote different sublattices, $i$ and $j$ indicate atomic positions, 
and $\bm S^{\mu}_i$ is the unit vector of the $i$ site in the $\mu$ sublattice. The Heisenberg
exchange parameters $J^{\mu\nu}_{\mathrm{ij}}$ are calculated by employing the Liechtenstein
formalism~\cite{liechtenstein1987local} within a self-consistent Green's-function method based on the multiple scattering theory within the density functional theory~\cite{Geilhufe2015}.
The crystalline structure information for the studied compounds obtained with the LCAO was used as input for electronic structure calculations by the Green's-function approach. According to our tests, both methods provide a very similar electronic structure for the systems under study. To estimate the Curie temperature $T_C$ we use the mean-field approximation for a multisublattice system~\cite{anderson1963theory,csacsiouglu2004first,yamada2018magnetic}, which is given by
\begin{equation}
\label{eqn:TcRPA}
T_C=\frac{2}{3k_B}J_{\mathrm{L}}^{\mu\nu},
\end{equation}
where $J_{\mathrm{L}}^{\mu\nu}$ is the largest eigenvalue of 
$J_{0}^{\mu\nu}=\sum_{j}J^{\mu\nu}_{0\mathrm{j}}$.

\section{Results and Discussion \label{sec:resanddis}}

We subdivide this section into three parts. First, we overview the ground-state electronic and magnetic properties of the SGSs (type I and type II) and HMMs based on Heusler compounds. In the second part, we analyze the tuning of type-II SGSs. In the third and final part, we discuss the exchange interactions, magnon dispersion, and Curie temperatures.

\subsection{Ground-state electronic and magnetic properties \label{sec:Ground}}

\begin{table*}[t]
\caption{\label{tab:StructureParam}
Optimized lattice constants $a_0$, sublattice and total magnetic moments, sum of the absolute values of the atomic spin magnetic moments $\sum_i |m_i|$, valence electron number Z$_T$, 
formation energy ($E_\text{form}$), convex hull distance energy ($\Delta E_\text{con}$), and calculated
and experimental Curie temperatures for 25 HMMs and SGSs. The $\Delta E_\text{con}$ and $E_\text{form}$
values are are taken from the Open Quantum Materials Database~\cite{saal2013materials}.} 
\begin{ruledtabular}
\begin{tabular}{@{}l*{11}{c}@{}}
Compound & $a_0$ & m$_X$ & m$_{X'}$ & m$_Y$ & m$_\text{total}$ & $\sum_i |m_i|$ & Z$_T$ & $E_\text{form}$ & 
$\Delta E_\text{con}$ & $T^{\mathrm{(MFA)}}_C$ & $T^{\mathrm{(exp)}}_C$\\
& ({\AA}) & ($\mu_B$) & ($\mu_B$) & ($\mu_B$) & ($\mu_B$) & ($\mu_B$) & & (eV/at.) & (eV/at.) & (K) & (K)\\
\hline
\multicolumn{11}{c}{Half-metallic magnets}\\
\hline
MnVTiAl & 6.11 & -2.54 & 2.60 & 0.91 & 1.00 & 6.08 & 19 & -0.172 & 0.188 & 963 & \\
MnVTiSi & 5.92 & -0.35 & 2.10 & 0.26 & 2.00 & 2.71 & 20 & -0.391 & 0.177 & 573 & \\
FeVTiAl & 6.06 & -0.78 & 2.42 & 0.45 & 2.00 & 3.75 & 20 & -0.247 & 0.117 & 685 & \\
FeVHfAl & 6.12 & -0.53 & 2.32 & 0.23 & 2.00 & 3.10 & 20 & -0.169 & 0.177 & 742 & \\
CoMnCrAs & 5.75 & 1.11 & -0.53 & 2.48 & 3.00 & 4.17 & 27 & -0.071 & 0.092 & 654 & \\
CoFeTiSi & 5.73 & 0.61 & 0.67 & -0.20 & 1.00 & 1.54 & 25 & -0.675 & 0.025 & 157 & \\
CoFeVSb & 5.99 & 1.08 & 1.20 & 0.78 & 3.00 & 3.12 & 27 & -0.016 & 0.198 & 308 & \\
CoFeCrSi & 5.61 & 1.04 & 0.22 & 1.86 & 3.00 & 3.24 & 27 & -0.293 & 0.075 & 517 & 
790~\cite{jin2016magnetism}\\
CoCoMnSi & 5.65 & 1.06 & 1.06 & 3.03 & 5.00 & 5.28 & 29 & -0.449 & 0.000 & 920 & 
985~\cite{kubler2007understanding}\\
\multicolumn{12}{c}{Spin-gapless semiconductors (type-I)}\\
\hline
MnCoMnAl & 5.73 & -2.01 & 0.99 & 3.03 & 2.00 & 6.04 & 26 & -0.271 & 0.035 & 1123 & 
720~\cite{ouardi2013realization}\\
CoMnCrSi & 5.63 & 0.92 & -0.96 & 2.07 & 2.00 & 3.98 & 26 & -0.334 & 0.065 & 589 & \\
CoFeTiSb & 6.08 & 1.06 & 1.33 & -0.33 & 2.00 & 2.78 & 26 & -0.325 & 0.190 & 476 & \\
CoFeTaGe & 5.94 & 1.07 & 1.14 & -0.26 & 2.00 & 2.52 & 26 & -0.248 & 0.127 & 453 & \\
CoFeCrAl & 5.69 & 0.97 & -0.71 & 1.84 & 2.00 & 3.62 & 26 & -0.199 & 0.108 & 421 & 
456~\cite{kharel2015magnetism}\\
\multicolumn{12}{c}{Spin-gapless semiconductors (type-II)}\\
\hline
MnCrNbAl & 6.07 & 1.36 & 2.49 & -0.74 & 3.00 & 4.71 & 21 & -0.181 & 0.033 & 624 & \\
MnCrTaAl & 6.06 & 1.30 & 2.44 & -0.63 & 3.00 & 4.49 & 21 & -0.208 & 0.030 & 637 & \\
FeVTiSi & 5.91 & 0.57 & 2.33 & 0.10 & 3.00 & 3.01 & 21 & -0.452 & 0.173 & 464 & \\
FeVHfSn & 6.40 & 0.30 & 2.63 & 0.12 & 3.00 & 3.10 & 21 & -0.148 & 0.139 & 705 & \\
FeVNbAl & 6.11 & 0.81 & 2.32 & -0.11 & 3.00 & 3.25 & 21 & -0.189 & 0.126 & 693 & \\
FeVTaAl & 6.10 & 0.79 & 2.32 & -0.11 & 3.00 & 3.23 & 21 & -0.213 & 0.096 & 681 & \\
FeCrTiAl & 5.96 & 0.48 & 3.08 & -0.44 & 3.00 & 4.14 & 21 & -0.310 & 0.036 & 560 & \\
FeCrHfAl & 6.15 & 0.27 & 3.18 & -0.31 & 3.00 & 3.90 & 21 & -0.236 & 0.060 & 568 & \\
RuCrHfAl & 6.30 & 0.07 & 3.44 & -0.32 & 3.00 & 4.02 & 21 & -0.458 & 0.064 & 669 & \\
OsCrHfAl & 6.31 & 0.12 & 3.37 & -0.33 & 3.00 & 3.99 & 21 & -0.392 & 0.064 & 428 & \\
CoOsCrAl & 5.86 & 0.86 & -0.39 & 1.66 & 2.00 & 3.04 & 26 & -0.248 & 0.062 & 369 & \\
\end{tabular}
\end{ruledtabular}
\end{table*}

The first step in our paper was to identify the Heusler compounds of potential interest. Then in the second step, we examined their electronic properties and we identified them as HMM or SGS. To carry out the first step we searched for type-I and type-II SGSs in the dataset of Gao \textit{et al.}~\cite{gao2019high} 
and calculated their electronic structure to identify candidates with large spin gaps. After selecting suitable materials we checked all of them in the Open Quantum Materials Database~\cite{saal2013materials}. Here we were interested in two energy quantities. The first one is the formation energy, $E_\text{form}$. This energy is the difference between the total energy of the \textit{XX'YZ} compound in the 
Heusler structure presented in \cref{fig:Heusler} and the sum of the energies of the isolated atoms of the chemical elements. This energy value should be negative in order to be able to grow the material in the Heusler structure. But this condition is not enough. The compound may prefer at this stoichiometry to grow in another structure or to separate in other phases (e.g. \textit{XY} and \textit{X'Z} binary compounds). For each stoichiometry, the phases with the minimum energy define the so-called convex hull. We decided to choose as our search filter a distance from the convex hull, $\Delta E_\text{con}$, less than 0.2 eV per atom because we think growing the compound in the Heusler structure as a metastable phase in the form of a thin film is possible since half-metallic CrAs in zinc-blende structure (space group \textit{F}$\overline{4}3$\textit{m}) with a hull distance of nearly 
0.3 eV/at. (see Supplemental Material of Ref. \cite{wang2017magnetic}) was stabilized on GaAs(001) by using molecular beam
epitaxy~\cite{mizuguchi2002epitaxial, mizuguchi2002epitaxialCrAs, akinaga2000material, akinaga2004zinc}.
Then for all the compounds which we identified to be of potential interest, we calculated the equilibrium lattice constant by minimizing the total energy and calculated the electronic structure. We have used the graphs presenting the total DOS versus the energy to identify HMM and SGS compounds (the DOS figures for all studied compounds are presented in the Supplemental Material). In \cref{tab:StructureParam}, we present the final 25 quaternary Heusler compounds (only CoCoMnSi is really a usual full-Heusler compound Co$_2$MnSi), 
which we found to have negative $E_\text{form}$, $\Delta E_\text{con}$ less than 0.2 eV per atom, and band structure compatible 
with a HMM or a SGS (type-I or type-II). Among the 25 studied compounds, only CoFeVSb and CoMnCrAs have 
small absolute values of $E_\text{form}$, close to zero, which may affect their stability. All other compounds 
present a $E_\text{form}$ absolute value quite high with CoFeTiSi being the most stable with a $E_\text{form}$ value of 
-0.675 eV per atom as it can be seen in \cref{tab:StructureParam}. With respect to the convex hull energy 
distances, the values in \cref{tab:StructureParam} are very encouraging. Especially almost all type-II SGS 
studied compounds present $\Delta E_\text{con}$ less than 0.1 eV per atom making them very promising to be 
grown in the form of thin films. Finally, we briefly comment on the equilibrium lattice constants 
$a_0$ presented also in \cref{tab:StructureParam}. The calculated values are between 5.6 and 6.4 {\AA} and 
there are a lot of HMM (type-I SGS) and type-II SGS combinations where the lattice parameters $a_0$ match. 
For example, the HMM MnVTiSi and type-II SGS FeVTiSi have lattice constants which differ less than 0.01 {\AA}.

The HMM or SGS character of the materials under study (see \cref{fig:SGS-HMM} for a schematic representation 
of the density of states) is compatible with the behavior of the total spin- magnetic moment. First, we focus on the HMM materials. For the ordered quaternary Heusler compounds, it is well known from Ref.~\cite{ozdougan2013slater} that the total spin magnetic moment in the unit cell $m_\text{total}$ (in 
units of $\mu_B$) versus the total number of valence electrons in the unit cell $Z_T$ follows a 
Slater-Pauling rule:
\begin{equation}
    m_\text{total} = Z_T - 18 
    \text{ or }
    m_\text{total} = Z_T - 24.
\end{equation}
This rule means that there are exactly 9 or 12 occupied minority-spin bands, respectively. As demonstrated in \cref{tab:StructureParam}, where we 
present also the total number of valence electrons Z$_T$, all \textit{XX'YZ} compounds where \textit{X'} is V or Cr fulfill the first variant of the rule while the rest of the compounds fulfill the second variant. In the first case there are 19, 20, or 21 valence electrons per unit cell while in the second case the number
of valence electrons in the unit cell is 25, 26, or 27. This behavior is clearly explained in Ref.~\cite{ozdougan2013slater}.
When \textit{X'} is V or Cr in the minority-spin band structure the triple degenerate at the $\Gamma$-point $t_{1u}$ states are high in energy and are unoccupied and thus there are in total nine occupied minority-spin states and the gap in the minority-spin band structure is formed between the occupied $t_{2g}$ 
and the unoccupied $t_{1u}$ states. When \textit{X'} is a heavier atom then the $t_{1u}$ states are located lower in energy, being fully occupied, and the gap in the minority-spin band structure is formed between these states and the empty double degenerate at the $\Gamma$-point $e_{u}$ states. Note that both the $e_{u}$ and $t_{1u}$ states obey the octahedral symmetry and not the tetrahedral symmetry of the lattice and thus are localized at the 4\textit{a} and 4\textit{b} sites occupied by the \textit{X} and \textit{X'} atoms.

In order to have a SGS material, the latter should have exactly 21 or 26 valence electrons per unit cell and thus a total spin magnetic moment of 3 $\mu_B$ or 2 $\mu_B$, respectively (note that in the case of 21 valence electrons the majority-spin (minority-spin) bands are now the spin-down (spin-up) bands and the Slater-Pauling rule is $m_\text{total} = 24-Z_T$, resulting in a positive value of the total spin magnetic moment). The origin of these two numbers, 21 and 26, has been extensively discussed in Ref.~\cite{ozdougan2013slater} and a schematic representation is given in Fig. 2 of this reference. To have a SGS the Fermi level should fall within gaps in both spin directions. In the case of 26 valence electron compounds, the situation is as in the usual HMM. In the minority-spin band structure, there are exactly 12 occupied bands. In the majority-spin band structure also the two $e_u$ states are occupied which are separated by a gap from the unoccupied antibonding $e_g$ and $t_{2g}$ states. In the case of the compounds with 21 valence electrons, the majority-spin band structure is similar to the minority-spin band structure of the HMM with exactly 12 occupied bands. In the minority-spin band structure, the $t_{1u}$ states are now empty, there are exactly nine occupied minority-spin bands, and there is a gap between the $t_{1u}$ states and the bonding $t_{2g}$ states which are just below them in energy. We remark in \cref{tab:StructureParam} that all five type-I SGS materials have 26 valence electrons, while all type-II SGSs with the exception of CoOsCrAl have 21 valence electrons.

We should also briefly discuss the spin magnetic moments in these compounds presented in
\cref{tab:StructureParam}. The total spin mag- netic moments are quite high for all studied compounds, being 2 
or 3 $\mu_B$. Only CoCoMnSi has a total spin magnetic moment of 5 $\mu_B$ and the HMM MnVTiAl and 
CoFeTISi of 1 $\mu_B$. These large values of the total spin magnetic moment stem from thelarge atomic spin magnetic moments of the transition-metal
atoms. Depending on the \textit{X}, \textit{X'} and \textit{Y} chemical elements,
the atomic spin magnetic moments at the various sites are
ferromagnetically or antiferromagnetically coupled, resulting
in ferrimagnetic compounds in most cases. As we stated above the \textit{X} and \textit{X'} atoms sit at the 4\textit{a} and 4\textit{b} sites, which are
the corners of a cube, being next-nearest (second) neighbors. The \textit{Y} and 
\textit{Z} atoms sit at the 4\textit{c} and 4\textit{d} 
sites at the center
of these cubes, being nearest (first) neighbors with the \textit{X} and \textit{X'} atoms. The \textit{Z} atoms are 
metalloids (also known as $sp$ elements) carrying negligible atomic spin magnetic moments;
for this reason, we do not show them in \cref{tab:StructureParam}. Thus, the \textit{Y} atom plays a crucial role, being the intermediary atom
between the \textit{X} and \textit{X'}. The late transition-metal atoms (Fe, Co,...) tend to have parallel spin magnetic moments when they are nearest neighbors, while the early transition-metal atoms (Mn, Cr,...) have the tendency to have antiparallel spin magnetic moments. We discuss the behavior of orientation of the atomic spin magnetic moments more in detail in the next section.

\begin{figure}[t]
\centering
\includegraphics[width=0.47\textwidth]{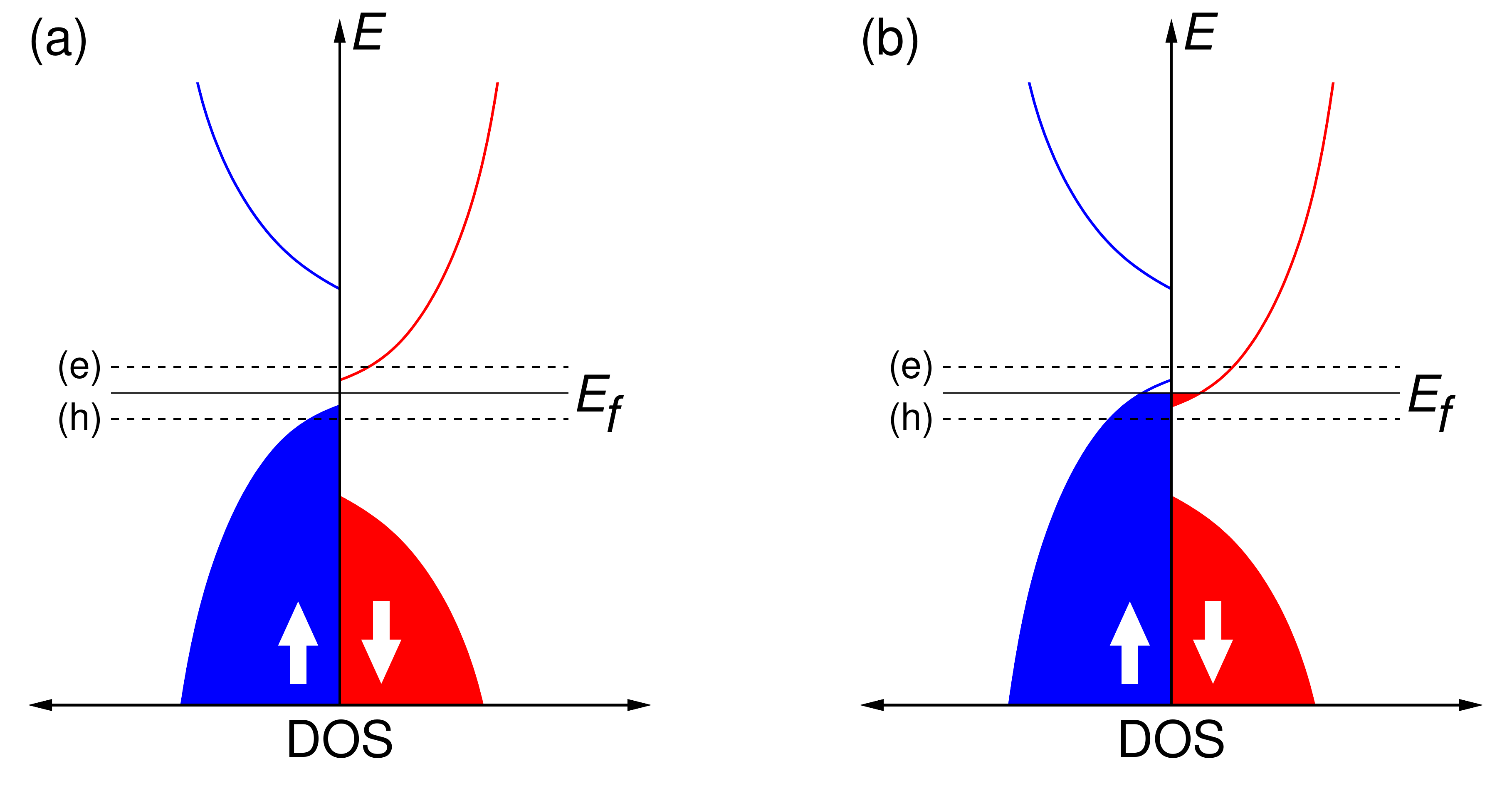}
\vspace{-0.2cm}
\caption{(a) Schematic representation of the density of states for a type-II SGS with a small gap between the majority- and minority- spin bands at the Fermi energy $E_F$ . (b) The same as (a) with a small overlap of bands of different spin channels. $E_F$ denotes the Fermi level, and the letters (e) and (h) represent electronlike and holelike behavior, respectively.}
\label{fig:SGS-real}
\end{figure}

\begin{figure*}[t]
\centering
\includegraphics[width=0.9\textwidth]{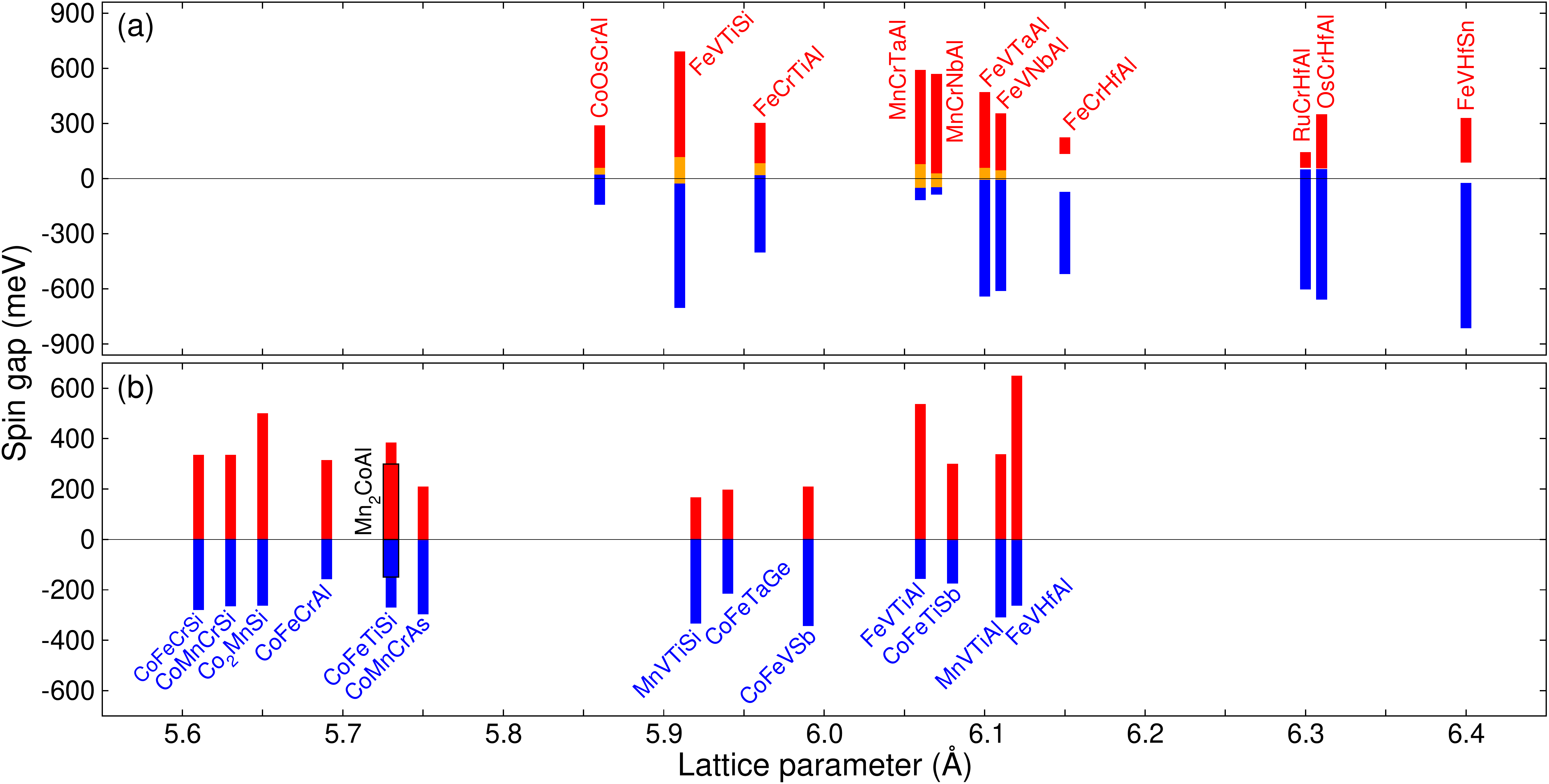}
\vspace{-0.1cm}
\caption{Comparison of the lattice parameters (a) for type-II SGSs and (b) for HMMs as well as type-I SGSs.The blue (black) and red (dark gray) bars illustrate the size of the gap below and above Fermi level $E_F$, respectively. The orange (light gray) or white bars represent the value of the overlap or the spin gap, respectively (see discussion in text). The Fermi level is located at 0 meV.}
\label{fig:SGS-HMM_comparison}
\end{figure*}

The most important quantity for the compounds under study is the width of the several gaps. First, we will start our discussion from the HMM and type-I SGS materials. In both cases as shown in \cref{fig:SGS-HMM} 
there is a gap in the minority-spin band structure and the Fermi level $E_F$ falls within this gap, splitting it into two parts, one below and one above $E_F$. In the majority-spin band structure, $E_F$ either intersects the bands (HMM case) or falls exactly within the zero-energy gap (type-I SGS). In the lower panel of \cref{fig:SGS-HMM_comparison}, we present for all HMMs and type-I SGSs the calculated spin-minority energy gaps, coloring with blue the part below $E_F$ and with red the part of the gap which is above $E_F$. 
The materials are ordered with ascending equilibrium lattice constant. For applications, we need materials with large energy gaps and with $E_F$ close to the center of the gap (comparable gaps below and above $E_F$) 
in order to minimize the effect of defects which usually induce states at the edges of the bands. We remark that all compounds possess gaps which are quite large (exceeding 0.4 eV) and in some cases like FeVHfAl they are close to 1 eV. Also for all compounds under study both parts of the minority-spin gap below (blue color) and above (red color) $E_F$ are sizable and thus are promising for the applications like magnetic tunnel diodes and transistors.

A more subtle case is the type-II SGS. Now we have a gap in both majority- and minority-spin band structures. In the ideal case, the maximum of the majority-spin valence band touches the minimum of the minority-spin conduction band as shown in \cref{fig:SGS-HMM}. In reality for all compounds under study, this ideal case does not occur. First, as shown in the left panel of \cref{fig:SGS-real} there can be a finite gap between the maximum of the majority-spin valence band and the minimum of the minority-spin conduction band. This is the case for the type-II SGS materials with the larger lattice constants: FeCrHfAl, RuCrHfAl, OsCrHfAl, and FeHfSn. In the upper panel of Fig. \cref{fig:SGS-HMM_comparison} we display the results for these compounds. The white space separating the blue and red regions is the gap between the majority-spin VBM and the minority-spin CBM. This is sizable in the case of FeCrHfAl and FeVHfSn, and almost vanishing for RuCrHfAl and OsCrHfAl. The blue bars mark the part of the gap which is located exclusively in the minority-spin band structure as shown in the left panel of \cref{fig:SGS-real} 
and with red bars we indicate the part of the gap which is located in the majority-spin band structure. The Fermi level is within the white region since we should have an integer number of occupied bands in both spin directions. In the case of RuCrHfAl and OsCrHfAl, the Fermi level intersects slightly the blue color and thus the valence majority-spin band structure but this is an artifact of the calculations due to numerical accuracy during the calculation of the density of states and this is easily confirmed if one extracts the band structure itself. If one tunes, as described in the next section, the position of the Fermi level, one can shift the Fermi level either within the majority-spin valence band, creating a hole surplus in the materials (the new position of the Fermi level is denoted with a dashed line and an "h" in the left panel of 
\cref{fig:SGS-real}), or within the minority-spin conduction band, creating a surplus of electrons (dashed line with "e" in the left panel of \cref{fig:SGS-real}).

In the rest of the type-II SGS compounds, there is an overlap between the majority-spin VBM and the minority- spin CBM as shown in the right panel of \cref{fig:SGS-real}. Now the Fermi level intersects both the majority-spin valence band and the minority-spin conduction band. This is clearly shown in the upper panel of \cref{fig:SGS-HMM_comparison} where the region of overlap for these compounds is denoted by an orange region and the $E_F$ for nearly all these compounds falls within the orange region. Below and above the orange region are the blue and red regions which denote the part of the energy gaps below and above the Fermi level which are located exclusively at the minority-spin and majority-spin band structures, respectively. A small shift of the Fermi level as discussed above can lead to a material with a hole or electron surplus which can be used as carriers in the material. There are materials like FeVTiSi, FeVTaAl, and FeVNbAl which present very large values of gaps both below and above the Fermi level and would be ideal for reconfigurable spintronic devices. Comparing the lattice constants, one observes in \cref{fig:SGS-HMM_comparison} that for realistic devices one has to use type-II SGSs with an overlap of the bands, because the type-II SGSs discussed in the above paragraph, which present a gap between the VBM and CBM, have very large lattice constants with respect to the HMMs.

\subsection{Tuning the type-II SGS \label{sec:Tune}}

\begin{figure}[t]
    \centering
    \includegraphics[width=0.45\textwidth]{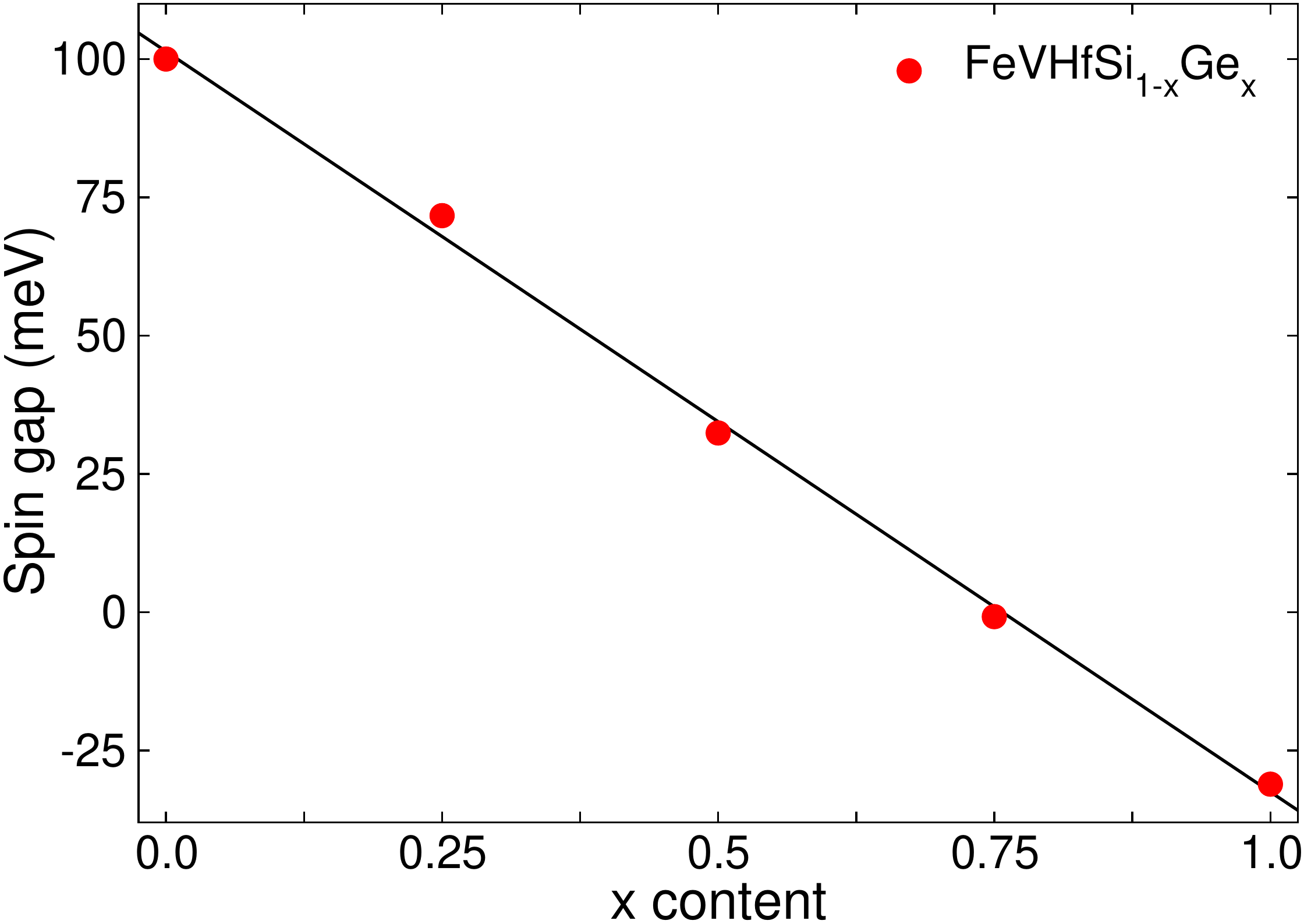}
    \caption{Indirect spin gap in FeVHfSi$_{1-x}$Ge$_x$ as a function of Ge concentration. 
    The black line displays the linear fit. For $x$=0.757 we get a perfect type-II SGS.}
    \label{fig:gap_tuning}
\end{figure}

To achieve the fabrication of the devices discussed in Sec. \ref{sec:Motiv}, one needs to use perfect 
type-II SGS. The maximum of the majority-spin valence band and the minimum of the minority-spin conduction band should be located exactly at the same energy position, which should be also the Fermi level. None of the compounds discussed above and presented in \cref{tab:StructureParam} and in \cref{fig:SGS-HMM_comparison} is 
a perfect type-II SGS. Thus, we should search for a way to tune the properties of these compounds. An obvious way to achieve that should be to start with two parent compounds presenting a spin gap (white region) and an overlap (orange region) in \cref{fig:SGS-HMM_comparison} and mix them.
Adding the correct fraction of each compound would lead to the disappearance of the overlap and to a perfect type-II SGS~(see \cref{fig:gap_tuning}).

But one has to be careful in choosing the two candidates for the mixture. These compounds should not differ in more than two elements and both elements have to be located in the same sublayer. So, for example, mixing FeVHfSi with FeVHfGe works (\cref{fig:gap_tuning}) while mixing FeCrHfAl with 
FeVTaAl does not. In the last example, the compounds differ only in the \textit{X'} and \textit{Y} element, 
but these two are located in different sublayers (see \cref{fig:Heusler}). In the 
material FeVHfSi$_{0.243}$Ge$_{0.757}$ the conduction band and the valence band 
would touch at $E_F$ (see \cref{fig:gap_tuning}). But please note that alloying can cause other undesirable side effects. In the case that particular states of different alloy components are close in energy, alloying can lead to a substantial band broadening. The band broadening depends also on the concentration. To avoid this side effect one can use alloy components, the states of which are separated in energy or are located far from the Fermi energy. In the latter case the band broadening affects the state far below the band gap area. Another possible effect of alloying is the change of the compound stoichiometry, which can also lead to the desired effect without band broadening the band edges. Furthermore, Heusler alloys can be doped with other elements. Hence, shifting the Fermi level to touch the minimum of the conduction band or the maximum of the valence band is possible.

We also checked if it is possible to achieve a band touching by adding strain or hydrostatic pressure. Compressing the samples by 5 GPa changes the lattice constant around 1\% but does not affect the electronic properties. {\c{S}}a{\c{s}}{\i}o{\u{g}}lu 
\textit{et al.} and Gavriliuk \textit{et al.} investigated the dependency of the Curie temperature on the applied pressure. In both cases $T_C$ is increasing with increasing 
pressure~\cite{sasioglu2005pressure, gavriliuk1996hyperfine}.
Shigeta \textit{et al.} analyzed the effect of pressure on the magnetic moment in 
Co$_2$TiSn and could not identify a change while applying pressure up to 1.27 GPa~\cite{shigeta2018pressure}. 
o investigate the effect of strain we built an eight-atom tetragonal unit cell ($a = b \neq c; \alpha = \beta = \gamma = \pi/2$)
and calculated the electronic structure when the $c$ axis was contracted or expanded while the volume of the cell stayed constant. So for the $a$ and $b$ axis, we followed the formula
\begin{gather}
    a = b = \sqrt{\frac{V}{c \cdot (1-x)}},
\end{gather}
where $x$ denotes the applied strain and $V$ stands for the volume of the cell.
This eventuates in a change of the electronic properties. Some bands are shifted to higher and some to lower energy. Thus, a general rule when the gap is closing could not be identified.

\subsection{Exchange interactions and Curie temperature \label{sec:JijandTc}}

For realistic applications of spintronic devices, the Curie temperature $T_C$ of the electrode
materials in tunnel junctions is extremely important. Materials with $T_C$ values much above 
room temperature are required. Most of the experimentally existing half-metallic Heusler compounds fulfill this requirement with $T_C$ values ranging from 300 to 1100 K. Compounds with highest 
reported $T_C$ values such as Co$_2$MnSi (985 K~\cite{kubler2007understanding}) and Co$_2$FeSi 
(1100 K~\cite{kubler2007understanding}) possess also large sublattice and thus total magnetic moments of 5 $\mu_B$ and 6 $\mu_B$, respectively. Extensive \textit{ab initio} calculations on multisublattice 
Heusler alloys have shown that there are several exchange interactions which coexist and are superimposed. Hence, a straightforward separation of the contributions of different mechanisms is not easy since DFT is not based on a model Hamiltonian approach and does not use a perturbative treatment. Exchange coupling in Heusler compounds, in which the total magnetic moment is localized on one sublattice (usually Mn-based compounds), is well understood on the basis of the Anderson \textit{s-d} mixing model~\cite{Anderson_sd_1,Anderson_sd_2,Anderson_sd_3,Anderson_sd_4}. 
It was shown that due to the large spatial separation of the Mn atoms in Heusler alloys ($d_{\mathrm{Mn-Mn}}> 4$\,{\AA}), 
the Mn 3$d$ states belonging to different atoms do not overlap considerably. Thus, an indirect exchange interaction between Mn atoms should play a crucial role in determining the magnetic state of the systems. However, the situation is different for the compounds studied here since the large part of the total magnetic moment is carried by two or three magnetic atoms with spatial separations of $2.5-3$ {\AA}. 
Therefore, the direct exchange coupling between the nearest magnetic atoms can dominate over the indirect one.

In order to simplify the discussion we can write the total magnetic exchange field acting on the sublattice $\mu$ as $J^{\mu}_{\mathrm{total}} \sim J^{\mu\nu}_{\mathrm{direct}} + J^{\mu\nu}_{\mathrm{indirect}}
+J^{\mu\mu}_{\mathrm{indirect}}$, where the first two terms represent the direct and indirect exchange couplings between different sublattices. The last term is intrasublattice indirect coupling. In compounds like Co$_2$MnSi and Mn$_2$CoAl in which the \textit{Y} sublattice carries a large magnetic moment the direct coupling provides the leading contribution to the total exchange coupling and determines the character of the magnetic state~\cite{sasiouglu2005exchange}. In most of the compounds considered, especially in type-II SGSs (see Table~\ref{tab:StructureParam}) the \textit{X} and \textit{X'} 
sublattices carry the magnetic moment. These sublattices have an interatomic distance 
$d_{\mathrm{X-X'}} \sim 3$ {\AA} and thus direct and indirect exchange coupling 
becomes important. It should be noted here that, in reality, the situation is not so simple and the exchange field acting on the sublattices should be determined from the solution of a matrix equation.

\begin{figure}[t]
\centering
\includegraphics[width=0.45\textwidth]{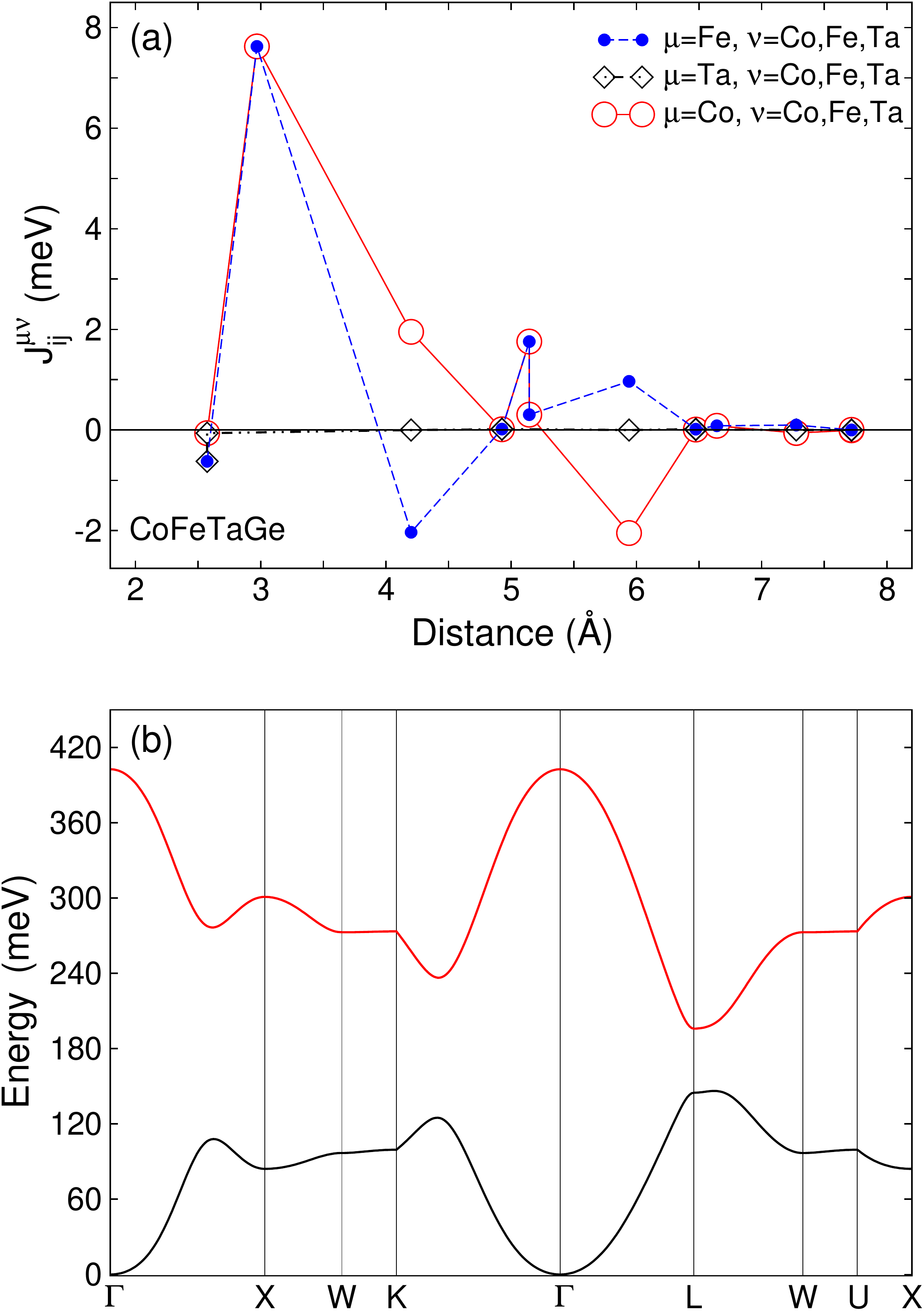}
\caption{(a) Intersublattice ($\mu \neq \nu$) and intrasublattice 
($\mu = \nu$) Heisenberg exchange parameters as a function of distance for type-I spin-gapless semiconductor CoFeTaGe. (b) Calculated magnon dispersion along the high-symmetry lines in the Brillouin zone for CoFeTaGe. The black curve represents the acoustic mode while red (dark gray) illustrates the optical branch.}
\label{fig:JijCoFeTaGe}
\end{figure}

\begin{figure}[t]
\centering
\includegraphics[width=0.45\textwidth]{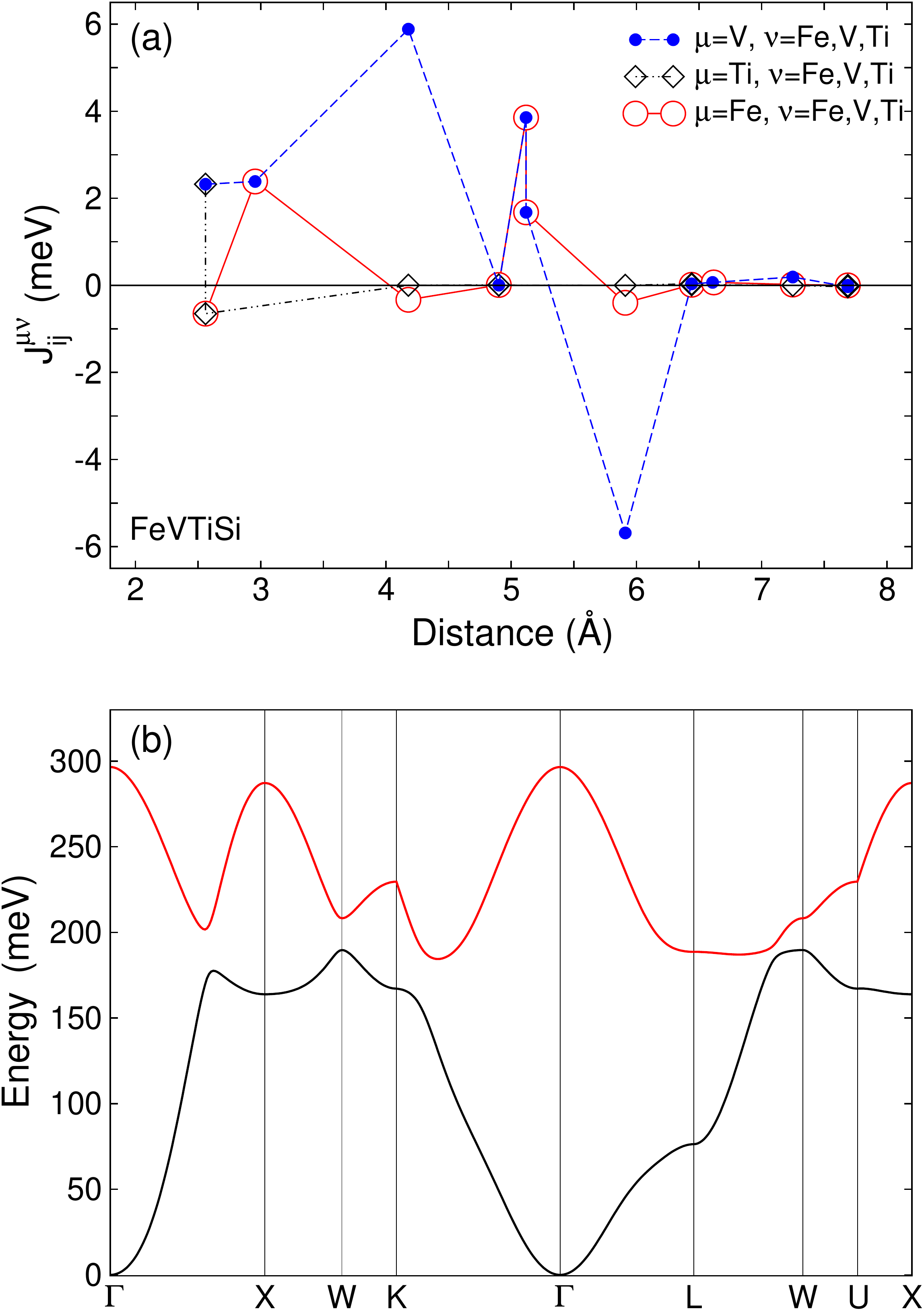}
\caption{((a) Intersublattice ($\mu \neq \nu$) and intrasublattice 
($\mu = \nu$) Heisenberg exchange parameters as a function of distance for type-II spin-gapless semiconductor FeVTiSi. (b) Calculated magnon dispersion along the high-symmetry lines in the Brillouin zone for FeVTiSi. The black curve represents the acoustic mode while red (dark gray) illustrates the optical branch.}
\label{fig:JijFeVTiSi}
\end{figure}

Due to the presence of a spin gap in both HMMs and SGSs the exchange interactions decay quickly with distance~\cite{rusz2006exchange, kudrnovsky2008electronic}. As representative of the type-I and type-II SGSs in Figs.~\ref{fig:JijCoFeTaGe} and \ref{fig:JijFeVTiSi} we present the calculated intrasublattice and intersublattice Heisenberg exchange parameters and corresponding magnon dispersion for CoFeTaGe (type-I SGS) and FeVTiSi (type-II SGS) compounds, respectively. As seen in both materials the intersublattice as well as the intrasublattice exchange parameters quickly decay with distance and for the interatomic separations larger than 8 {\AA} all parameters vanish. 
In both com- pounds, the Co and Fe (Fe and V) sublattices form a cubic cell. In the case of CoFeTaGe, the Co and Fe sublattices possess similar magnetic moments of about 1.1 $\mu_B$, while the Ta atom has a small induced magnetic moment of -0.26 $\mu_B$, which couples antiferromagnetically to the Co and Fe sublattices. As seen in Fig. \hyperref[fig:JijCoFeTaGe]{7(a)} the intersublattice Fe-Ta as well as Co-Ta interactions are almost negligible despite very short interatomic distance of $d_{\mathrm{Fe-Ta}}=2.57$ {\AA}.
This means that the Ta sublattice is more or less decoupled from the rest of the system.

In CoFeTaGe the strongest interaction takes place between
the Fe and Co sublattices and it quickly decays with distance, i.e., from $J_1^{\mathrm{Fe-Co}} \sim 8$ meV to 
$J_2^{\mathrm{Fe-Co}}\sim 2$ meV and $J_3^{\mathrm{Fe-Co}}$ becomes zero. On the other hand, the intrasublattice Fe-Fe and Co-Co exchange interactions behave very differently, i.e., they show Ruderman-Kittel-Kasuya-Yosida-type oscillations with strong damping, however with different sign and more or less with the same amplitude. Thus, their contributions into the total exchange coupling almost cancel each other and only Fe-Co intersublattice exchange interactions play a decisive role in determining ground-state and finite-temperature properties of the type-I SGS compound CoFeTaGe.

The situation is a bit different for the type-II SGS FeVTiSi compound, in which V sublattice carries a large magnetic moment
of 2.33 $\mu_B$, while Fe and Ti sublattices have rel- atively small magnetic moments of 0.57 $\mu_B$ and 0.1 $\mu_B$, respectively. Due to different sublattice magnetic moments the patterns of calculated exchange parameters presented in Fig. \hyperref[fig:JijFeVTiSi]{8(a)} are also different than in the CoFeTaGe compound.
In FeVTiSi the Ti sublattice couples ferromagnetically to
the Fe and V sublattices due to strong ferromagnetic V-Ti
intersublattice exchange interaction, while the Fe-Ti interaction is antiferromagnetic but its strength is one-third of
the V-Ti interaction and thus the overall contribution turns
out to be ferromagnetic. The Fe and V sublattices interact ferromagnetically with $J_1^{\mathrm{Fe-V}} > J_2^{\mathrm{Fe-V}}$ 
and $J_2^{\mathrm{Fe-V}}$ splits into two due to different exchange paths along the [111] direction [see 
Fig. \hyperref[fig:JijFeVTiSi]{8(a)}]. The strongest interaction in FeVTiSi takes place between nearest- and next-nearest-neighbor V atoms, which have opposite sign and similar strength. Note that each V atom has 12 nearest-neighbor and 6 next-nearest neighbor V atoms. Furthermore, the intrasublattice Fe-Fe interactions are antiferromagnetic but negligibly small. Moreover, all exchange parameters quickly decay with distance and become zero after 8 {\AA}.
Note also that in all other type-II SGSs, except CoOsCrAl, the \textit{X'} sublattice (V or Cr atoms) carries a 
large magnetic moment (see Table~\ref{tab:StructureParam}) and, as a result, the calculated patterns of intrasublattice exchange parameters (results not shown) are similar to the FeVTiSi case. In most of the type-II SGSs the \textit{Y} sublattice couples antiferromagnetically to the \textit{X} and \textit{X'} sublattices. However, this coupling is weak due to the small magnetic moment of atoms in the \textit{Y} sublattice.

As mentioned in the preceding section the ferrimagnetic ground state in most of the considered compounds (20 out of 25) can be qualitatively accounted for on the basis that half-filled shells tend to yield a strong trend toward antiferromagnetism. As seen in Table~\ref{tab:StructureParam}, when the \textit{Y} sublattice is occupied by the Cr (Mn) atom and the \textit{X'} sublattice is occupied by Mn or Fe (Os) the coupling between these sublattices is antiferromagnetic since both Cr and Mn atoms possess half-filled 3d shells and Fe (Os) is close to half filling. Most of the materials satisfy either one or both conditions.

In Figs.\,\hyperref[fig:JijCoFeTaGe]{7(b)} and \hyperref[fig:JijFeVTiSi]{8(b)} we present the magnon dispersion along the high-symmetry lines in the Brillouin zone for CoFeTaGe and FeVTiSi, respectively. Note that for both compounds the induced small magnetic moments on Ta and Ti atoms are not treated as independent degrees of freedom in magnon dispersion calculations and thus we have only two branches. The acoustic branches in both materials are typical for magnets with short-range interactions, where nearest-neighbor and next-nearest-neighbor intersublattice and intrasublattice exchange interactions dominate, and do not yield any magnetic instabilities. Magnetic instabilities can occur if the acoustic magnon modes have very low (close to zero) or negative energies in some parts of the Brillouin zone but this is not the case for any of the studied compounds. Around the $\Gamma$ point
point the energy-dispersion curves show a quadratic behavior with spin-wave stiffness constants of $D=224$ meV {\AA}$^2$ for CoFeTaGe and $D=314$ meV {\AA}$^2$ 
for FeVTiSi. These values are comparable to the typical values of transition-metal ferromagnets which usually range between 300 and 600 meV {\AA}$^2$.

The optical magnon branch, which corresponds to the out-of-phase precession of magnetic moments in \textit{X} and \textit{X'} sublattices, has a strong dispersion in both compounds. As the magnetic moments in \textit{X} and \textit{X'} sublattices in CoFeTaGe have similar values the optical branch looks like a mirror image of the acoustic branch 
[see Fig.\,\hyperref[fig:JijCoFeTaGe]{7(b)}].

Calculated exchange parameters are used to estimate the Curie temperature $T_C$ within the multisublattice mean-field approximation [see eq.~\ref{eqn:TcRPA}]. The obtained $T_C$ 
values are presented in \cref{tab:StructureParam}. For comparison, available experimental data are also included. As seen for all compounds except CoFeTiSi the estimated $T_C$ 
values are above room temperature, ranging from 308 K to 1123 K. Our mean-field 
estimation of $T_C$ for Co$_2$MnSi and CoFeCrAl is in reasonable agreement with available experimental data. However, $T_C$ for Mn$_2$CoAl is overestimated, which can be attributed to the mean-field approach. As in the mean-field approach spin fluctuations are assumed to be small and the spin-flip Stoner excitations are neglected, it gives the upper bound for $T_C$ values, however in materials with large coordination number (fcc lattice) and with long-range exchange interactions the mean-field $T_C$ values are close to the ones obtained with random- phase approximation and classical Monte Carlo methods. Of course, this is not the case for the
Mn$_2$CoAl compound, which possesses very large nearest-neighbor intersublattice Mn-Mn and Mn-Co exchange interactions and, as a result, mean-field considerably overestimates the $T_C$
by 50\%, while the Monte Carlo method results in a $T_C$ value of 770\,K~\cite{jakobsson2015first}.
Note that in HMMs and type-I SGSs the presence of spin gap around the Fermi energy prevents spin-flip transitions. Thus, Stoner excitations do not play an important role in the thermodynamics of these materials.

On the other hand, underestimation of $T_C$ by about 35\% in CoFeCrSi compound can be attributed to the long-wavelength approximation in linear response theory, which underestimates exchange parameters in materials with small magnetic moments like fcc Ni, which has been discussed extensively in the literature by several authors~\cite{Bruno,antropov2003exchange,antropov2006calculation,katsnelson2004magnetic,dederichs1984ground,jacobsson2017parameterisation}. 
In the case of the CoFeCrSi compound, the Fe atom has a small magnetic moment of 0.22 $\mu_B$ and thus the long-wavelength approximation in linear response theory is expected to underestimate the intersublattice Fe-Co as well as the Fe-Cr exchange parameters, and as a consequence we obtain a small $T_C$ value of 517 K compared 
to the experimental value of 790 K. 
Due to the long-wavelength approximation our estimated $T_C$ values might be smaller than the experimental values when these materials are grown since most of the considered compounds have one or two transition-metal sublattices with small magnetic moments.

\begin{figure}[t]
\centering
\includegraphics[width=0.43\textwidth]{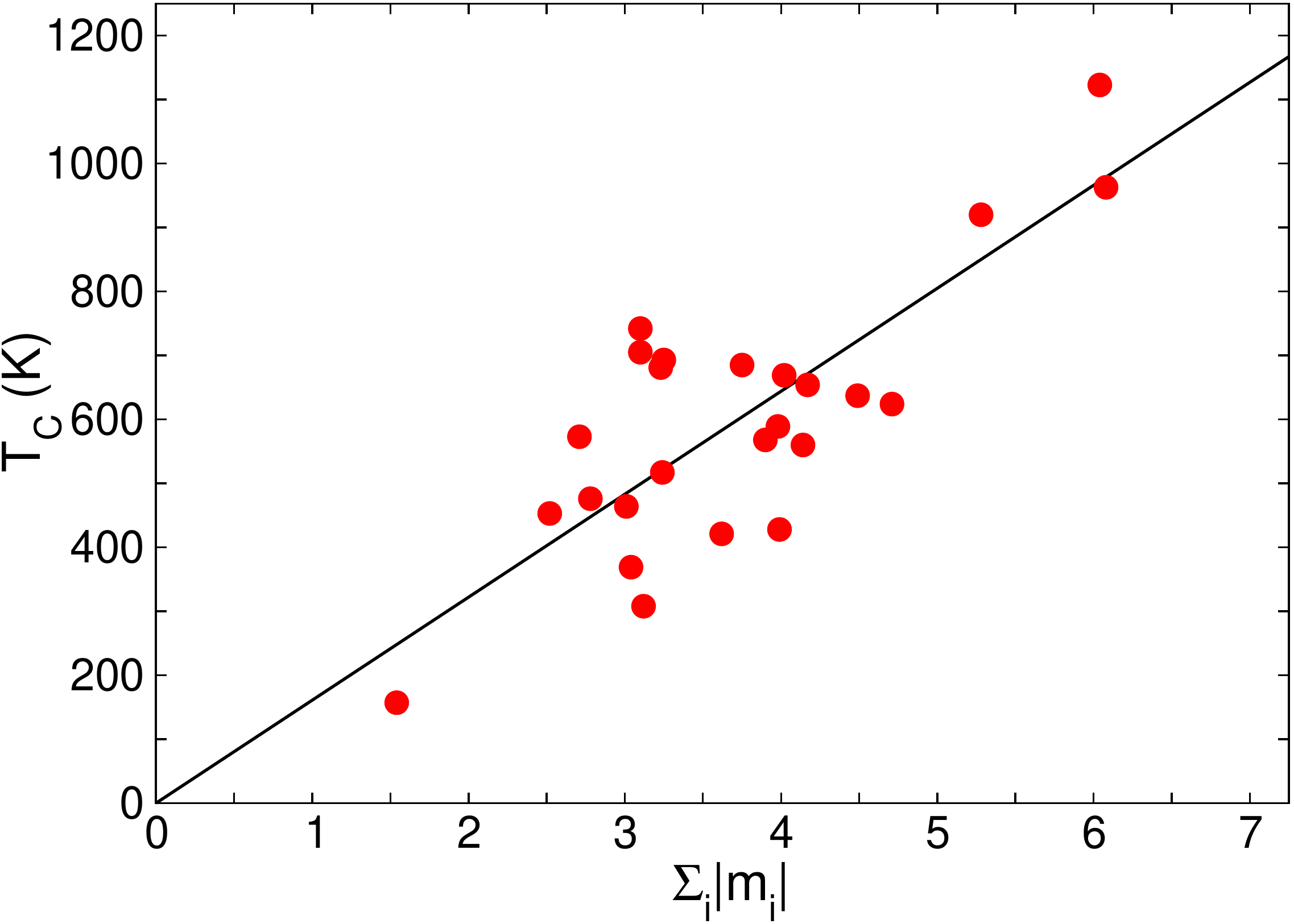}
\caption{The dependence of the calculated Curie temperatures on the sum of the absolute values of the 
sublattice magnetic moments $m_T^{\mathrm{abs}}$ ($m_T^{\mathrm{abs}}=\sum_i |m_i|$) presented in 
\cref{tab:StructureParam}. The solid line displays the linear fit $y=161 \cdot x$.}
\label{fig:Tc-M}
\end{figure}

Finally, we would like to comment on the semiempirical
relation between calculated $T_C$ values
and the sum of the absolute values of the sublattice magnetic moments $m_T^{\mathrm{abs}}=\sum_i |m_i|$ 
which are presented in \cref{tab:StructureParam}. 
The relation between $T_C$ and $m_T^{\mathrm{abs}}$ is presented in \cref{fig:Tc-M}. As seen the 
$T_C$ increases almost linearly, $T_C \sim 161 \cdot m_T^{\mathrm{abs}}$, with increasing 
$m_T^{\mathrm{abs}}$, and materials with largest $m_T^{\mathrm{abs}}$ values like Mn$_2$CoAl 
and Co$_2$MnSi possess also the highest $T_C$ values. Most of the compounds have $m_T^{\mathrm{abs}}$ 
values in between 2.5 $\mu_B$ and 5 $\mu_B$ and thus moderate Curie temperatures. 
Deviations from the linear behavior
can be traced back to the sublattice magnetic moments and
thus the pattern of exchange interactions. In materials like FeVHfAl, 80\% of the $m_T^{\mathrm{abs}}$ 
is carried by the V sublattice and thus intrasublattice V-V exchange interactions play a decisive role in the formation of $T_C$ rather than intrasublattice exchange interactions. Of course, no such general rule exists since also compounds like CoFeVSb with similar sublattice magnetic moments show strong deviation.

\section{Conclusion \label{sec:concl}}

Spintronics is a rapidly developing area of nanoelectronics. The emergence of new concepts like reconfigurable magnetic tunnel diodes and transistors requires the design of materials with novel functionalities. For that purpose, Heusler compounds are a preferential choice to identify such materials. In the present paper, we searched suitable half-metallic magnets and spin-gapless semiconductors among the family of ordered quaternary Heusler compounds with the chemical formula \textit{XX'YZ} 
to realize reconfigurable magnetic tunnel diodes and transistors. We managed to identify 25 compounds which combine HMM or SGS properties with negative formation energies and small convex hull energy distances so that they can be grown experimentally.

Following the identification of the compounds of interest, we employed state-of-the-art 
\textit{ab initio} electronic band-structure calculations to determine their lattice constant, the spin magnetic moments, and their electronic structure. The total spin magnetic moment of all compounds exhibits a Slater-Pauling behavior and the ones being SGS have either 21 or 26 valence electrons per unit cell as expected for SGSs. Among the ones that are SGSs, there are five of the so-called type I which possess a gap in the minority-spin band structure and a zero gap in the majority-spin band structure. The other 11 SGS compounds are of type II, presenting gaps in both spin directions. None of these 11 compounds is a perfect SGS but as we show suitable mixing of two parent compounds leads to the tuning of their electronic properties and the appearance of perfect SGS type-II characteristics (the maximum of the majority-spin valence band and the minimum of the minority-spin conduction band touch exactly at the Fermi level). All compounds present large values of atomic spin magnetic moments and the calculated exchange constants are short-range stabilizing the magnetic state. We calculated the Curie temperatures for all 25 compounds and found them to be well above room temperature. 

We expect that our results will pave the way for experimentalists to fabricate magnetic tunnel diodes and transistors by combining suitable HMM and SGS quaternary Heusler compounds.

\begin{acknowledgments}
E.\c{S}	and I.M. gratefully acknowledge funding provided by the European Union (EFRE) and by Deutsche Forschungsgemeinschaft SFB Grant No. TRR 227.
\end{acknowledgments}

\bibliographystyle{apsrev4-2}
\bibliography{bibliography}

\end{document}


\title{Supplemental Material}

\author{T. Aull$^{1}$}
\author{E. \c{S}a\c{s}{\i}o\u{g}lu$^{1}$}
\author{I. V. Maznichenko$^{1}$}
\author{S. Ostanin$^{1}$}
\author{A. Ernst$^{2,3}$}
\author{I. Mertig$^{1,2}$}
\author{I. Galanakis$^{4}$}

\affiliation{$^{1}$Institute of Physics, Martin Luther University Halle-Wittenberg, D-06120 Halle (Saale), Germany \\
$^{2}$Max Planck Institute of Microstructure Physics, Weinberg 2, D-06120 Halle (Saale), Germany\\
$^{3}$Institute for Theoretical Physics, Johannes Kepler University Linz, Altenberger Straße 69, A-4040 Linz, Austria\\
$^{4}$Department of Materials Science, School of Natural Sciences, University of Patras, GR-26504 Patra, Greece}

\maketitle

\thispagestyle{fancy}

In the main text we discuss electronic and magnetic properties of in total 25 half-metallic 
magnets (HMMs) and spin-gapless semiconductors (SGSs) for realization of reconfigurable spintronic 
devices. In this supplementary part we provide the total density of states (DOS) for all 25 compounds.

\begin{figure}
    \centering
    \includegraphics[width=\textwidth]{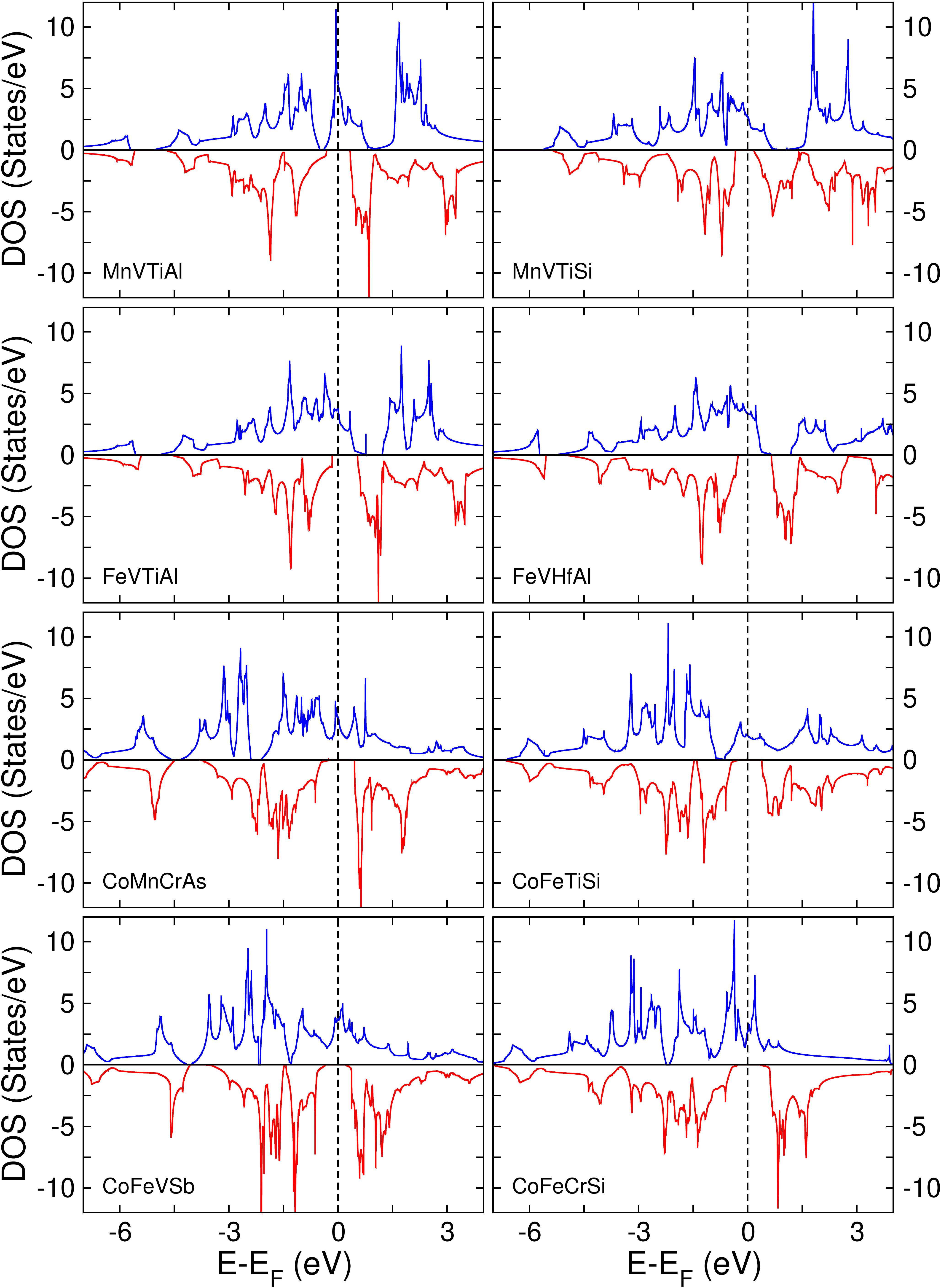}
    \label{fig:DOS_HMM}
\end{figure}

\begin{figure}
    \centering
    \includegraphics[width=\textwidth]{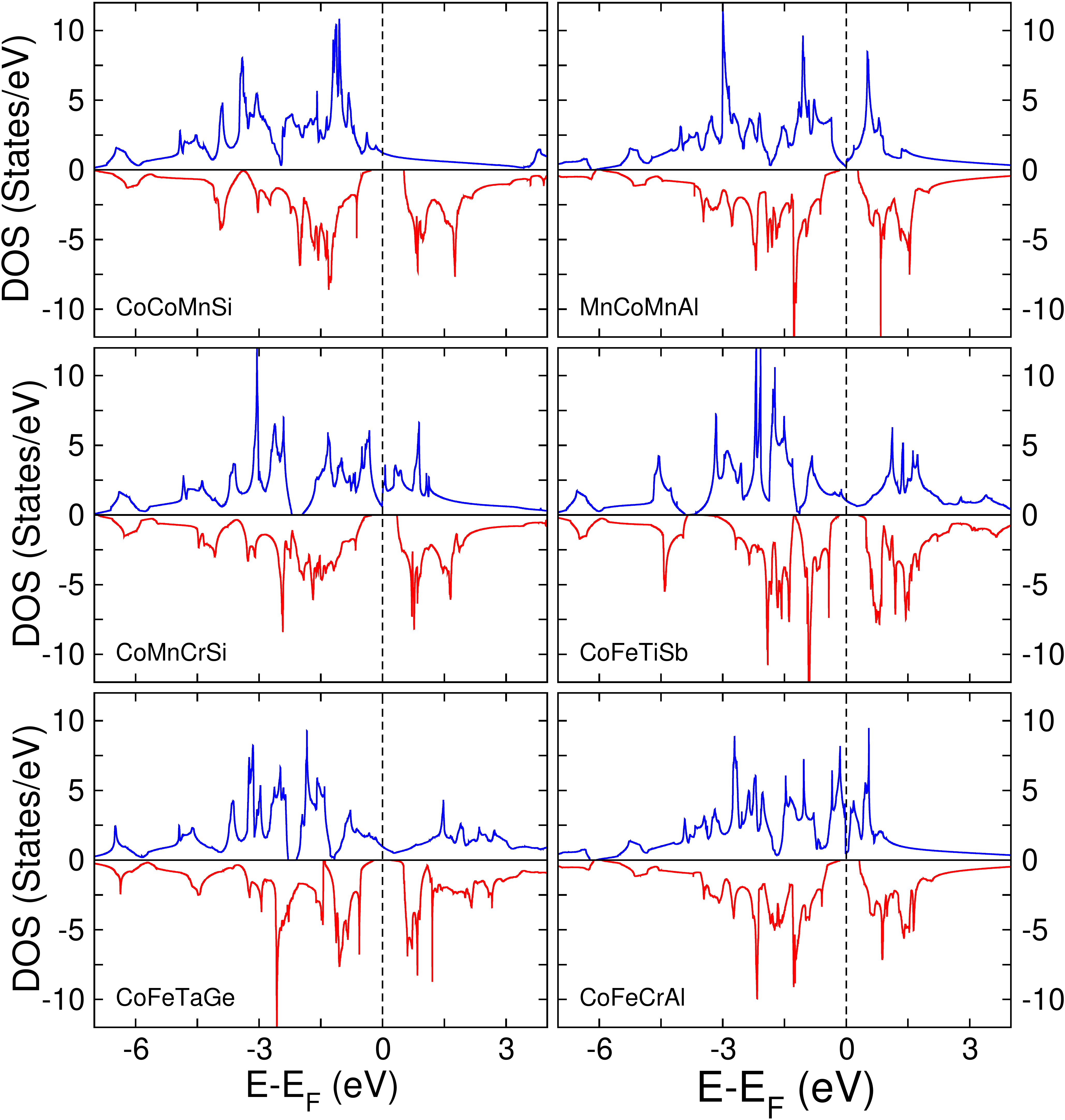}
    \caption{Total density of states for the HMMs and  type-I SGSs. 
    With blue and red color we display the majority spin bands and minority spin bands, respectively.}
    \label{fig:DOS_HMM_SGSI}
\end{figure}

\begin{figure}
    \centering
    \includegraphics[width=\textwidth]{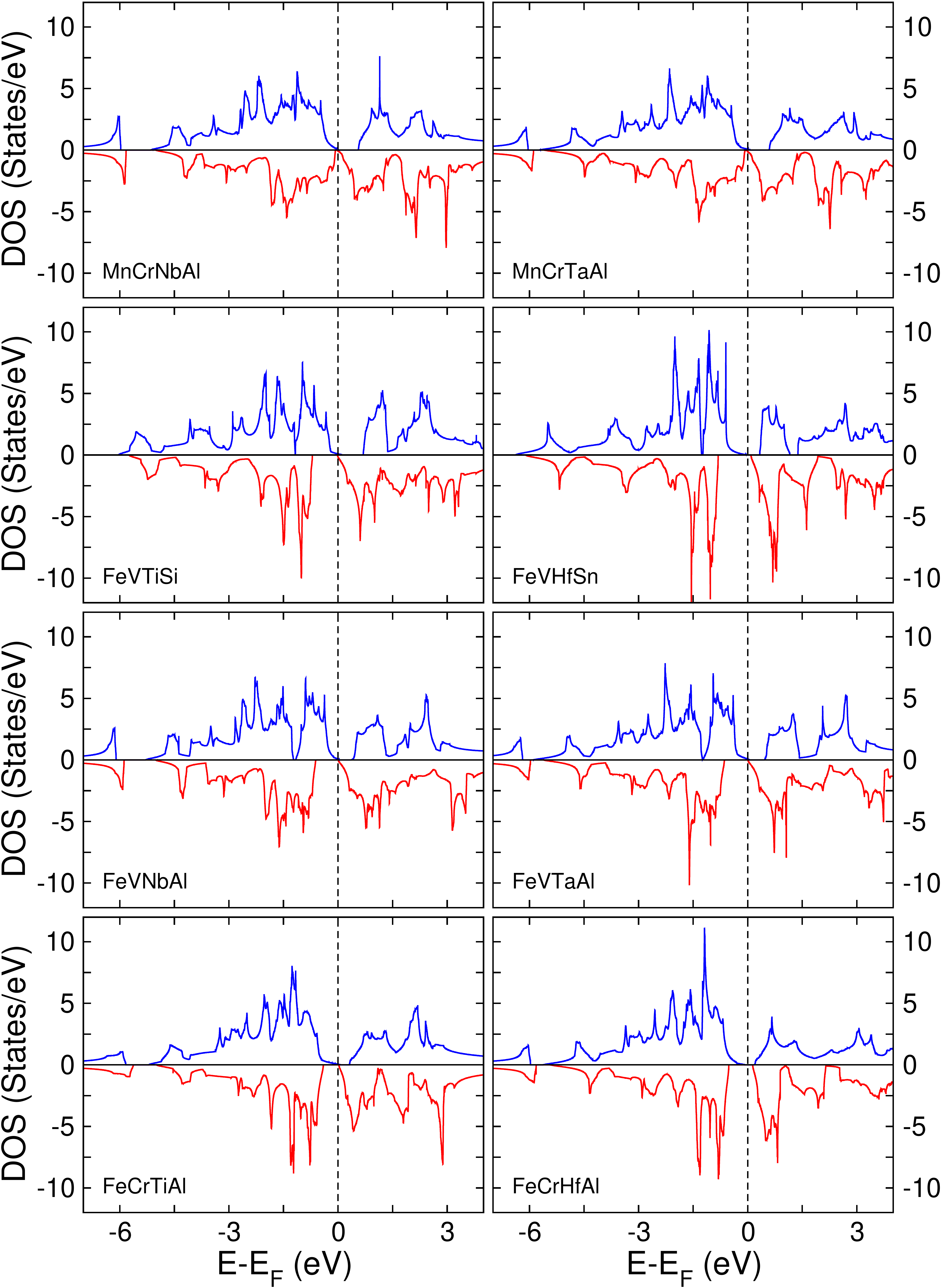}
    \label{fig:DOS_SGSII_1}
\end{figure}

\begin{figure}
    \centering
    \includegraphics[width=\textwidth]{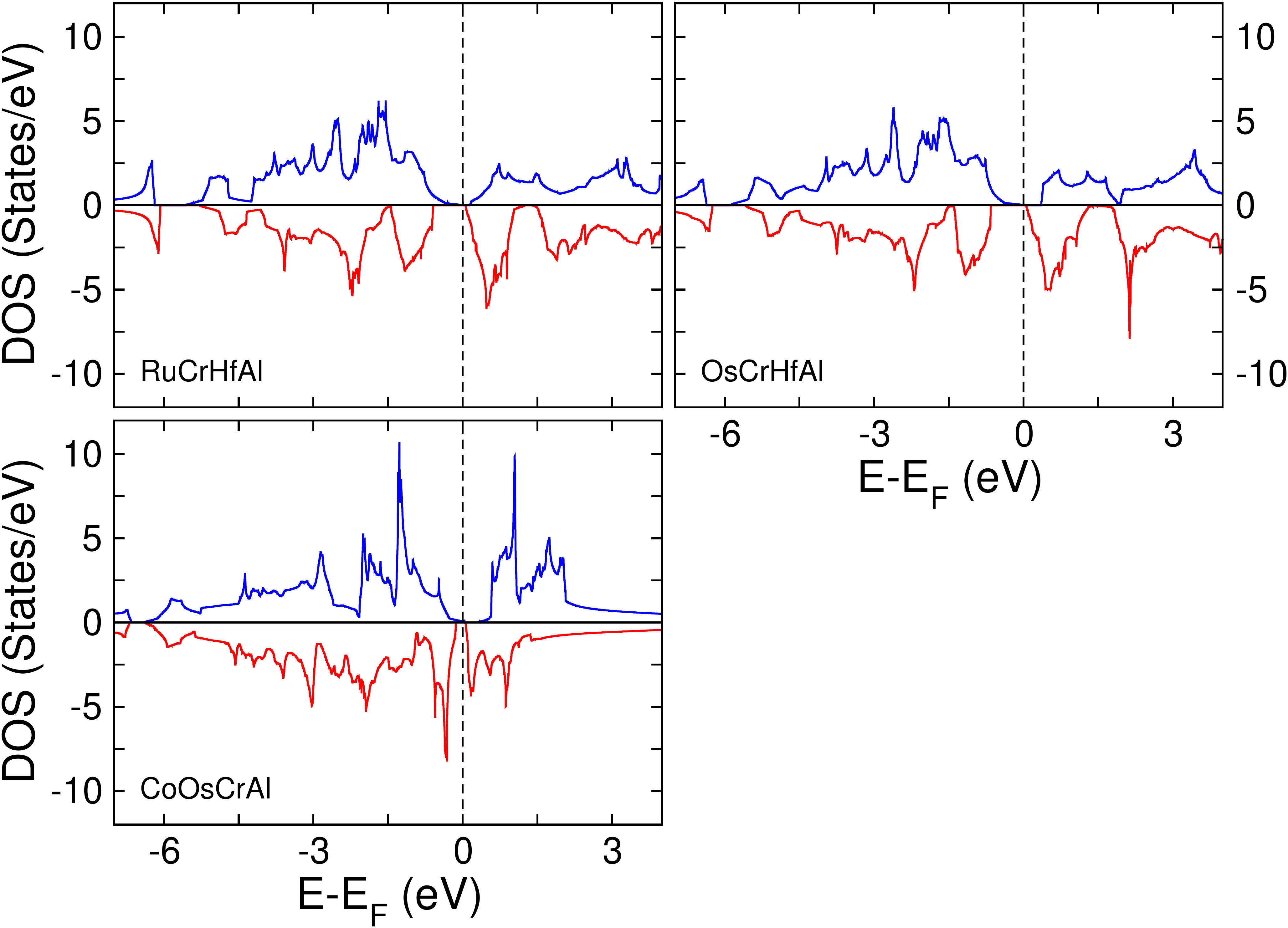}
    \caption{Total density of states for the type-II SGSs. 
    With blue and red color we display the majority spin bands and minority spin bands, respectively.}
    \label{fig:DOS_SGSII_2}
\end{figure}